\documentclass[aps,prd,nofootinbib,showpacs,superscriptaddress, preprint]{revtex4-1}
\usepackage{graphicx}
\usepackage{epsfig}
\usepackage{amsmath}
\usepackage{amsfonts}
\usepackage{float}
\usepackage{amssymb}
\usepackage{color}
\usepackage[bottom]{footmisc}
\usepackage{array,multirow,makecell}
\usepackage{amsfonts}
\def\lsim{\raise0.3ex\hbox{$\;<$\kern-0.75em\raise-1.1ex\hbox{$\sim\;$}}}
\def\gsim{\raise0.3ex\hbox{$\;>$\kern-0.75em\raise-1.1ex\hbox{$\sim\;$}}}
\def\be{\begin{equation}}
\def\ee{\end{equation}}
\def\bea{\begin{eqnarray}}
\def\eea{\end{eqnarray}}

\begin{document}

\title{Common origin of modified chaotic inflation, non thermal dark matter and Dirac neutrino mass}
\author{ Debasish Borah}
\email{dborah@iitg.ac.in}
\affiliation{Department of Physics, Indian Institute of Technology Guwahati, Assam 781039, India}
\author{Dibyendu Nanda}
\email{dibyendu.nanda@iitg.ac.in}
\affiliation{Department of Physics, Indian Institute of Technology Guwahati, Assam 781039, India}
\author{Abhijit Kumar Saha}
\email{aks@prl.res.in}
\affiliation{Theoretical Physics Division, Physical Research Laboratory, Ahmedabad 380009, India}
\begin{abstract}
We propose a minimal extension of the standard model of particle physics to accommodate cosmic inflation, dark matter and light neutrino masses. While the inflationary phase is obtained from a modified chaotic inflation scenario, consistent with latest cosmology data, the dark matter particle is a fermion singlet which remains out of equilibrium in the early universe. The scalar field which revives the chaotic inflation scenario by suitable modification also assists in generating tiny couplings of dark matter with its mother particle, naturally realizing the non-thermal or freeze-in type dark matter scenario. Interestingly, the same assisting scalar field also helps in realizing tiny Yukawa couplings required to generate sub-eV Dirac neutrino mass from neutrino couplings to the standard model like Higgs field. The minimality as well as providing a unified solution to all three problems keep the model predictive at experiments spanning out to all frontiers.
\end{abstract}
\maketitle


\section{Introduction}
Several experimental evidences from cosmological and astrophysical experiments strongly favour
the presence of large amount of non-baryonic form of matter dubbed as Dark Matter (DM), in the present universe \cite{Tanabashi:2018oca}. However, mysteries surrounding the nature and properties of DM are yet to be resolved by the experiments operating at cosmic, intensity as well as energy frontiers. It is well known
that in order to accommodate DM candidate in particle physics, extension of the standard model (SM) of particle physics is required. Depending on the strength of the interactions of DM with the SM particles,
different types of DM models have been proposed. Among them, the weakly interacting massive
particle (WIMP) paradigm has been the most widely studied dark matter scenario for last few decades. In this framework, a DM candidate 
typically with electroweak scale mass and interaction rate similar to 
electroweak interactions can give rise to the correct DM relic 
abundance \cite{Aghanim:2018eyx}, a remarkable coincidence often referred to as the \textit{WIMP 
Miracle}. Now, if such type of particles whose interactions are of the order of 
electroweak interactions really exist then we should expect their signatures in 
various DM direct detection experiments where the recoil energies of detector 
nuclei scattered by DM particles are being measured. However, the null results at direct detection experiments \cite{Akerib:2015rjg,Aprile:2017iyp,Cui:2017nnn,Tan:2016zwf} have brought the several implementations of the WIMP scenario into tension. This has also resulted in an increased amount of interest in scenarios beyond the 
thermal WIMP paradigm where the interaction strength of DM particle with SM particles can be much 
lower than electroweak interaction i.e.\,\,DM may be more 
feebly interacting than in the thermal WIMP paradigm. One of the viable 
alternatives of WIMP paradigm, which may
be a possible reason behind null results at various direct detection
experiments, is to consider the non-thermal origin of DM \cite{Hall:2009bx}. For a recent review of such feebly interacting (or freeze-in) massive particle (FIMP) DM, please see \cite{Bernal:2017kxu}. In the FIMP scenario, DM candidate does not thermalise with the SM particles in the
early universe due to it's feeble interaction strength and the initial abundance of DM
is assumed to be zero. At some later stage, DM can be produced non thermally from decay or annihilation of other particles thermally present in the universe.

Apart from the above mentioned DM problem which brings cosmology and particle physics close to each other in the pursuit of finding a solution, another issue faced by the standard hot big bang cosmology is related to the observed isotropy of the cosmic microwave background (CMB), also known as the horizon problem. This can not be explained in the standard cosmology where the universe remains radiation dominated throughout the early stages. In order to solve this, the presence of an rapid accelerated expansion phase in the early universe, called inflation \cite{Guth:1980zm,Starobinsky:1980te} was proposed. Originally proposed to solve the horizon, flatness and unwanted relic problem in cosmology \cite{Guth:1980zm, Linde:1981mu}, the inflationary paradigm is also supported by the adiabatic and scale invariant perturbations observed in the CMB \cite{Komatsu:2010fb, Akrami:2018odb}. It turns out that, one can accommodate such an early inflationary phase within different particle physics models where one or more scalar fields play the role of inflaton. Chaotic inflation \cite{Linde:1983gd,Martin:2013tda} models were one of the earliest and simplest scenarios that used power law potentials like $m^2\phi^2$ with a scalar field $\phi$. It predicts specific values for inflationary parameters like the spectral index $n_s\sim 0.967$, tensor-to-scalar ratio $r\sim 0.133$ for number of e-folds $N_e=60$. However Planck 2018 results \cite{Akrami:2018odb} strongly disfavour this simple model due to its large prediction of $r$. In order to continue using such simple power law type inflationary scenario, one then has to modify the simplest chaotic inflation potential. There exists several proposals in literature where such attempts have been made.  For 
example, radiative corrections to the inflationary potential \cite{NeferSenoguz:2008nn,Enqvist:2013eua,Ballesteros:2015noa}, non-minimal coupling of the inflaton with gravity \cite{Pallis:2014cda,Gumjudpai:2016ioy,Tenkanen:2017jih} or logarithimic
mass correction of the inflaton \cite{Kasuya:2018cxo, Dong:2018aak} could bring down the value of $r$ within the Planck limit as well as keeping $n_s$ within the
allowed range. Another interesting proposal exists, where the inflation sector is extended by an additional scalar field
 \cite{Harigaya:2015pea,Saha:2016ozn}. The mere interaction between the inflaton and the additional scalar field can revive
the model successfully by reducing the magnitude of $r$ below the Planck 2018 limit \cite{Akrami:2018odb}. It is to be noted that the presence of
linear or cubic terms of the inflaton in the Lagrangian can destroy the required flatness for having successful inflation in the simple chaotic inflationary scenarios. This can be ensured by imposing one discrete symmetry $Z_2$ under which $\phi$ transforms non trivially. Now, after the end 
of inflation, reheating of the universe is extremely essential so that the production of radiation and other matter fields can occur. This also sets the initial condition for the standard big-bang cosmology. The energy transfer of inflaton to relativistic matter fields can be realised through non perturbative preheating process or perturbative decay of inflaton. In particular, it was shown in \cite{Felder:1999pv} that thermalisation of the universe through instant preheating could be one simple and elegant scenario for $Z_2$-odd chaotic inflation model. For a recent work on $Z_2$-odd inflaton field where reheating occurs from annihilation instead of decay, please see \cite{Choubey:2017hsq, Borah:2018rca}. 

Although the fundamental origins of inflation and DM could be disconnected to each other, it is very well motivating to study them in a common framework. In fact there have been proposals where a single field can play the role of inflaton as well as DM \cite{Kofman:1994rk, Kofman:1997yn, Liddle:2006qz, Cardenas:2007xh, Panotopoulos:2007ri, Liddle:2008bm, Bose:2009kc, Lerner:2009xg, Okada:2010jd, DeSantiago:2011qb, Lerner:2011ge, delaMacorra:2012sb, Khoze:2013uia,  Kahlhoefer:2015jma, Bastero-Gil:2015lga, Tenkanen:2016twd, Choubey:2017hsq, Heurtier:2017nwl, Hooper:2018buz,Daido:2017tbr,Daido:2017wwb, Borah:2018rca, Almeida:2018oid, Manso:2018cba, Choi:2019osi}. Motivated by such common frameworks, here we study a scenario where inflation and non-thermal DM can find a common origin. However, unlike a similar proposal \cite{Tenkanen:2016twd} where non-thermal DM field acted as inflaton, here we consider a fermionic non-thermal DM which is produced dominantly from the decay of a scalar field that was present in the thermal bath of early universe. The decay is assisted by another scalar field which plays a non-trivial role in modifying the minimal chaotic inflation scenario as required by the latest cosmology data mentioned earlier. Thus, although the inflaton and DM fields are not the same, yet they are non-trivially connected by another scalar field which assists in successful inflation and DM production. The presence of DM sector 
also opens up the possibility of perturbative
inflaton decay at the end of inflation.

In order to make our framework more minimal and predictive, we also attempt to address the origin of light neutrino masses. Non-zero neutrino mass and large leptonic mixing are well established facts by now \cite{Tanabashi:2018oca} while their origin remains unknown as the SM can not explain them. Apart from neutrino oscillation experiments, cosmology experiments like Planck also constrain neutrino sector by putting an upper bound on the sum of absolute neutrino masses $\sum \lvert m_i \rvert < 0.12$ eV \cite{Aghanim:2018eyx}. While the nature of light neutrinos: Dirac or Majorana, remains undetermined at oscillation experiments, popular seesaw models like \cite{Minkowski:1977sc, GellMann:1980vs, Mohapatra:1979ia, Schechter:1980gr}, proposed to account for neutrino masses predict Majorana neutrinos. However, experiments looking for neutrinoless double beta decay ($0\nu \beta \beta$), a promising signature of Majorana neutrinos, have not yet found any positive results. Though this does not necessarily rule out the Majorana nature, yet it is motivating to study the possibility of light Dirac neutrinos. This has led to  
several proposals that attempt to generate tiny Dirac neutrino masses in a variety of ways \cite{Babu:1988yq, Peltoniemi:1992ss, Chulia:2016ngi, Aranda:2013gga, Chen:2015jta, Ma:2015mjd, Reig:2016ewy, Wang:2016lve, Wang:2017mcy, Wang:2006jy, Gabriel:2006ns, Davidson:2009ha, Davidson:2010sf, Bonilla:2016zef, Farzan:2012sa, Bonilla:2016diq, Ma:2016mwh, Ma:2017kgb, Borah:2016lrl, Borah:2016zbd, Borah:2016hqn, Borah:2017leo, CentellesChulia:2017koy, Bonilla:2017ekt, Memenga:2013vc, Borah:2017dmk, CentellesChulia:2018gwr, CentellesChulia:2018bkz, Han:2018zcn, Borah:2018gjk, Borah:2018nvu}. In a recent work \cite{Borah:2018gjk}, a common origin of light Dirac neutrinos and non-thermal DM was proposed where tiny couplings involved in non-thermal DM and Dirac neutrino mass had common source from higher dimensional operators. Here we extend that idea to incorporate inflation as well \footnote{See \cite{Allahverdi:2007wt, Mazumdar:2012qk, Mazumdar:2010sa, Kohri:2009ka, Liddle:2008bm, Rodrigues:2018jpv, Kazanas:2004kv} for earlier attempts in linking neutrino and dark matter with inflation.}, within a modified chaotic inflation scenario. We extend the SM by three additional scalar fields and a fermion DM field with suitable discrete symmetries in order to keep the unwanted terms away. Since light neutrinos are of Dirac-type, three right handed neutrinos are present by default. While one of the scalars is the inflaton field and one is the mother particle for DM, the third one assists in inflation as well as DM production. We find that correct DM and neutrino phenomenology can be successfully reproduced in the model while the inflationary parameters predicted by the model remain allowed from Planck 2018 data.

This paper is organised as follows. In section \ref{sec1}, we present our model and the corresponding particle spectra, including light Dirac neutrinos. In section \ref{sec2}, we discuss the details of inflation in our model followed by the details of non-thermal fermion DM in section \ref{sec3}. We finally conclude in section \ref{sec4}.

\section{The Model}
\label{sec1}
In this section, we discuss our model, its particle content, additional symmetries and the Lagrangian of the new fields. As mentioned in the introduction, we extend the SM with two gauge singlet real scalars ($\phi$ and $\chi$), one gauge singlet complex scalar
field ($\eta$), one vector like fermion ($\psi$)
and three right neutrinos ($\nu_R$). All the new fields considered in the model are singlets under the SM gauge symmetry. We impose two discrete symmetries $Z_4\times  Z_4^\prime$
in addition to the SM gauge symmetry, in order to achieve the desired terms in the Lagrangian \footnote{{To keep our discussion minimal, we have adopted such discrete symmetries. UV completion can be achieved by suitable gauge symmetries, for example Abelian gauge extensions \cite{Langacker:2008yv}. A recent work where $B-L$ gauge symmetry leading to light Dirac neutrinos and a residual $Z_2 \times Z^{\prime}_2$ symmetry can be found in \cite{Nanda:2019nqy}.}}. The charge assignments of the new fields under the discrete symmetries are shown in table \ref{tab:charge_assignments}. The SM fields other than leptons  have trivial charges $+1$ under both the discrete symmetries. We also consider an unbroken global lepton number symmetry $U(1)_L$ under which SM leptons as well as $\nu_R, \psi$ have unit charges. This ensures the absence of Majorana mass terms for neutrinos via higher dimensional operators, leading to a purely Dirac nature of light neutrinos.
\begin{table}[h]
\begin{center}
    \begin{tabular}{| l | l | l | l | l | l | l |}
    \hline
     & $l_L$ & $\nu_R$ & $\psi$ & $\chi$ & $\eta$ & $\phi$  \\ \hline
$Z_4$  & $1$ & -1 & 1 & -1 & 1 & -1   \\ \hline
$Z_4^\prime$  & $i$ & $i$ & -1 & 1 &  $i$ & -1  \\ \hline
    \end{tabular}
\end{center}
\caption{Charge assignments of the new fields present in the model under the discrete symmetries.}
\label{tab:charge_assignments}
\end{table}

The scalar Lagrangian as followed from the charge assignments is provided by
\begin{align}
V(\phi,\chi,\eta)=&\frac{1}{2}m^2\phi^2+\frac{\lambda_{\phi}}{4} \phi^4-\frac{c_1}{4}\Big(\chi^2-\frac{v_\chi^2}{2}\Big)\phi^2+\frac{\lambda_\chi}{4} \Big(\chi^2-\frac{v_\chi^2}{2}\Big)^2\nonumber \\
&+m_\eta^2|\eta|^2+
\lambda_\eta|\eta|^4
+\frac{\lambda_{\phi\eta}}{2}\phi^2|\eta|^2+\frac{\lambda_{\chi\eta}}{2}|\eta|^2\Big(\chi^2-\frac{v_\chi^2}{2}\Big)\nonumber\\
&+\lambda_H \Big(|H|^2-\frac{v^2}{2}\Big)^2+\lambda_{H\eta}|\eta|^2\Big(|H|^2-\frac{v^2}{2}\Big)\nonumber\\
&+\frac{\lambda_{\chi H}}{2}\Big(\chi^2-\frac{v_\chi^2}{2}\Big)\Big(|H|^2-\frac{v^2}{2}\Big) \nonumber \\
&+\frac{\lambda_{\phi H}}{2}\phi^2 \Big(|H|^2-\frac{v^2}{2}\Big)+ \left (\frac{\lambda_{R}}{2} \phi\eta \eta \chi+~{\rm h.c.} \right)~,
\label{eq:PotTot}
\end{align}
where we identify the SM Higgs doublet as $H$ and $v, v_{\chi}$ are the vacuum expectation values (VEV) of the neutral component of $H$ and $\chi$ respectively.

The relevant part of the fermionic Lagrangian consistent with the charge assignments of the fields
is given by
\begin{align}
\mathcal{L_F}\supset & m_\psi\bar{\psi}{\psi}+\Big(\frac{\xi_{\chi\psi}\chi^2}{M_P}
+\frac{\xi_{\phi\psi}\phi^2}{M_P}+\frac{\xi_{\eta\psi}|\eta|^2}{M_P}\Big)\bar{\psi}{\psi}\\
&+\left( \frac{\xi\eta \chi\bar{\psi}\nu_R}{M_P}+{\rm~h.c.~} \right)+\left ( \frac{y\chi \bar{l_L} \tilde{H} \nu_R}{M_P}+{\rm~h.c.~} \right).
\end{align}
Here $l_L$ denotes the usual lepton doublet of SM and $M_P$ is the Planck mass. The bare mass term for vector like fermion $\psi$ is the only renormalisable term involving $\psi$ while all other terms arise only at dimension five level or higher. The discrete symmetries prevent coupling between SM lepton doublet and right handed neutrinos at tree level which would require Dirac neutrino Yukawa to be fine tuned at the level of $\mathcal{O}(10^{-12})$ or even smaller. The discrete symmetries also help in preventing the Majorana mass term of right handed neutrinos, which is needed to ensure the pure Dirac nature of light neutrinos. After $\chi, H$ acquire non-zero VEV's, the light neutrino masses arise as 
\begin{equation}
M_{\nu} = y \frac{v v_{\chi}}{2M_P}.
\end{equation}
For $v \approx 10^2$ GeV, $v_{\chi} \approx 10^9$ GeV, it is possible to generate $M_{\nu} \approx 0.1$ eV for Yukawa couplings $y \approx 10^{-2}$. The light neutrino mixing will arise from the structure of $y$ in the flavour basis, which is not restricted by the symmetries of the model and hence it is possible to fit it with observed mixing \cite{Tanabashi:2018oca} without having any consequence for DM and inflation sector as we discuss below. Some of the higher dimensional terms involving $\psi$ can also generate a new contribution to $\psi$ mass after one or more of the scalar fields acquire non-zero VEV. However, since $\psi$ already has a bare mass term it is always possible to adjust its mass at a suitable value by adjustment of different relative contributions.


\section{Inflation}
\label{sec2}
\begin{figure}[h]
 \centering
 \includegraphics[height=5cm,width=8cm]{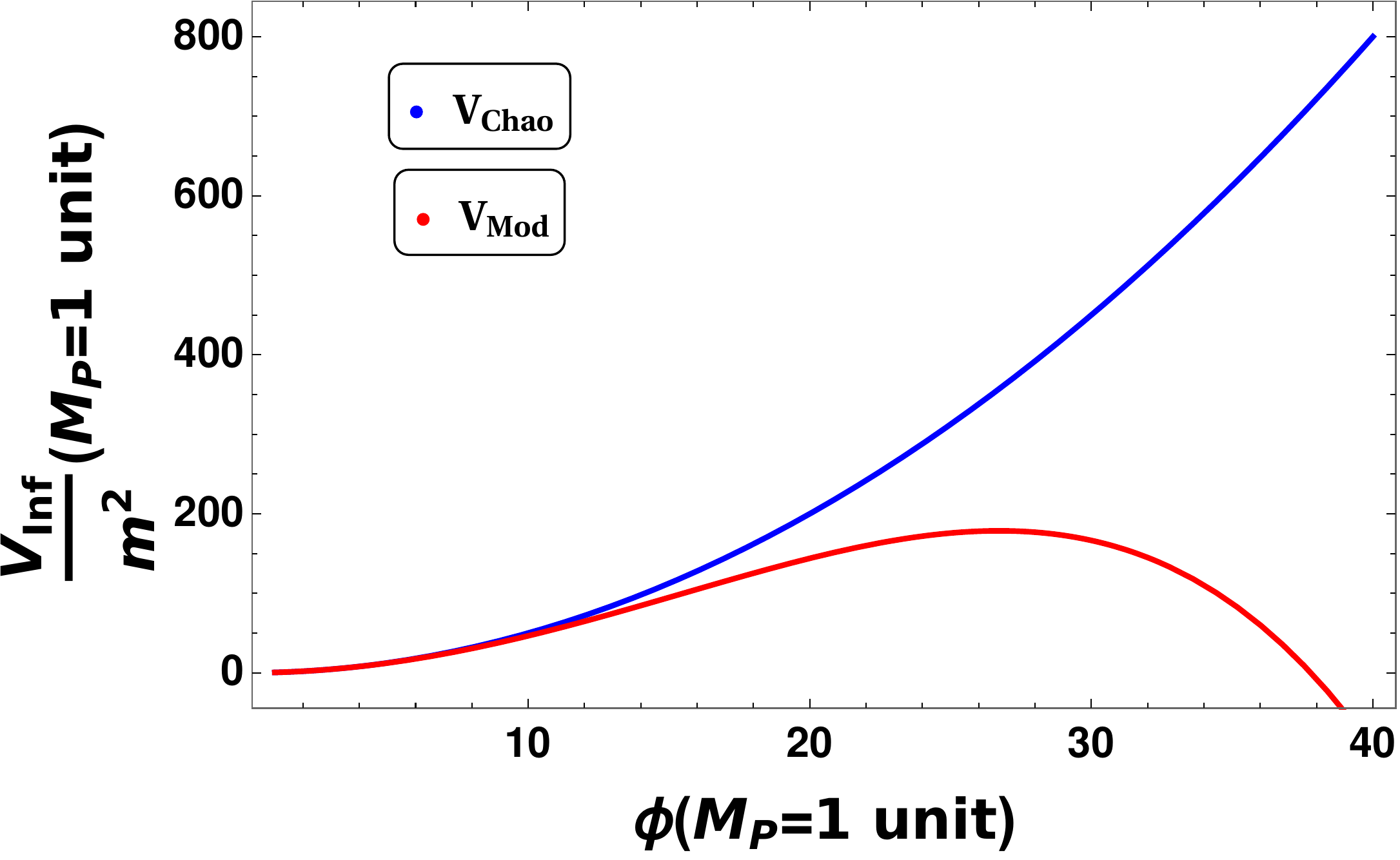}
  \includegraphics[height=5.5cm,width=8cm]{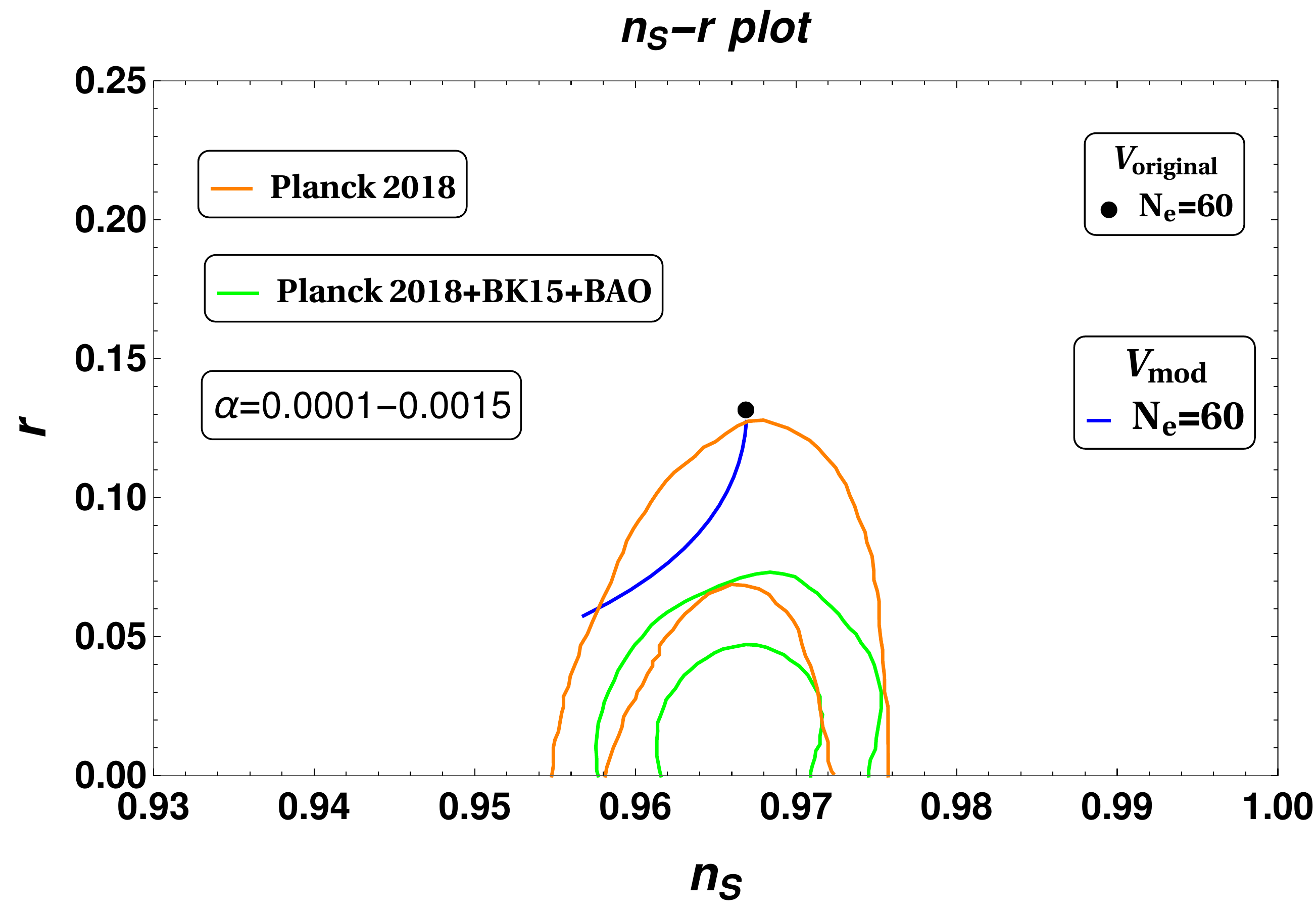}
 \caption{[Left] Sketch of $V_{\rm Inf}^{\rm eff}$ given in equation (\ref{eq:InfPeff}) for $\alpha=0.0005$ (red) and $\alpha=0$ (blue). [Right] $n_s$-$r$ contour line for  $N_e=60$ as obtained from our proposed model. For comparison
purpose we also show the prediction of the original chaotic inflation model. Furthermore we also include Planck TT, TE, EE +lowE+lensing and Planck TT, TE, EE +lowE+lensing+ BK15+BAO $1\sigma$ and $2\sigma$ allowed contours separately.}
 \label{fig:InfPo}
\end{figure}

\noindent Here we discuss the dynamics of inflation and its predictions in detail. The inflation is governed by the following potential,
\begin{align}
V(\phi,\chi)=\frac{m^2\phi^2}{2}-\frac{c_1}{4}\Big(\chi^2-\frac{v_\chi^2}{2}\Big)\phi^2+ \frac{\lambda_\chi}{4}\Big(\chi^2-\frac{v_\chi^2}{2}\Big)^2,\label{InfP}
\end{align}
where we identify $\phi$ as the inflaton and $\chi$ is the assisting field. We also consider the coupling coefficients $c_1$ and $\lambda_\chi$ to be real and positive. The global minimum of $\chi$ field is denoted by $v_\chi$ which we assume to be much smaller than $M_P$. We also ignore higher order terms of $\phi$ {\it e.g.} $\phi^4$ by considering the associated coupling coefficients negligibly small. At the beginning of inflation 
the $\chi$ field acquires a negative mass squared ($\sim c_1\phi^2$) of order $\mathcal{O}(H_{\rm Inf}^2)$ where $H_{\rm Inf}$ is the Hubble parameter during inflation. Thus the $\chi$ field is expected to be driven quickly towards its inflaton field dependent local minimum given by,
\begin{align}
\langle\chi^2\rangle_{\rm Inf}=\frac{v_\chi^2}{2}+\frac{c_1}{2\lambda_\chi}\phi^2~\simeq~ \frac{c_1}{2\lambda_\chi}\phi^2,
\label{chiV}
\end{align}
where we assume $v_\chi^2\ll \frac{c_1}{\lambda_\chi}\phi^2 $.
Around $\chi=\langle\chi\rangle_{\rm Inf}$, the effective mass
squared of the $\chi$ field is positive and obtained as
\begin{align}
m_{\chi}^2\Big|_{\chi=\langle\chi\rangle_{\rm Inf}}=c_1\phi^2+\lambda_\chi v_\chi^2~\simeq~ c_1\phi^2.\label{mChi}
\end{align}
This turns out to be bigger than the $H_{\textrm{Inf}}^2\simeq \frac{m^2\phi^2}{6 M_P^2}$ (with suitable choices of $c_1$ and $\lambda_\chi$ and super-Planckian $\phi$ during inflation) and hence the $\chi$ field is expected to be stabilised at $\langle\chi\rangle_{\rm Inf}$ with negligible fluctuations \cite{Dong:2010in,Evans:2015mta,McAllister:2014mpa,Buchmuller:2015oma}. Thus inflation occurs along $\phi$ field direction with the $\chi$ field stabilised at $\langle\chi\rangle_{\rm Inf}$.
One can obtain the effective inflationary potential by integrating out the heavier field $\chi$ \cite{Dong:2010in,Evans:2015mta,McAllister:2014mpa,Buchmuller:2015oma} (by replacing equation (\ref{chiV}) into equation (\ref{InfP})) which is given by,
\begin{align}
V_{\rm Inf}^{\rm eff}~=~&\frac{1}{2} m^2\phi^2-\frac{c_1^2}{16\lambda_\chi}\phi^4 \nonumber, \\
=~&\frac{m^2\phi^2}{2}\Big(1-\alpha\phi^2\Big),\label{eq:InfPeff}
\end{align}
where we write $\alpha=\frac{c_1^2}{8\lambda_\chi m^2}$. From here onwards, for making the analysis simple we shall work with $M_P=1$ unit. The parameter $\alpha$ determines the amount of deformation of the modified chaotic potential from the minimal chaotic inflation scenario. In left panel of figure \ref{fig:InfPo},
we plot the effective inflationary potential $V_{\rm Inf}^{\rm eff}$ ({normalised by $m^2$}) for $\alpha=0$ (blue) and $\alpha=0.0007$ (red). As we can see, non zero $\alpha$, associated with the presence of $\chi$ field, makes the inflationary potential flatter than the minimal one ($\alpha=0$). The flattening starts to occur near $\phi\sim 10$ in $M_P=1$ unit for $\alpha=0.0007$ while before that the potential merges with the original chaotic inflation potential. Although flattened, the potential becomes unbounded from below at large value of $\phi$ ({\it e.g.} after the maximum at $\phi\sim 30$ for $\alpha=0.0007$) as seen from left panel of figure \ref{fig:InfPo}. Then in order to have successful inflation  we must make an important assumption that inflaton always stays below
the maximum of the potential towards the flat part. However the possibility of tunnelling of the inflaton from its minimum to the unstable part of the potential  still remains.   
In that case it is essential to confirm the metastability of the minimum of $\phi$ by calculating the corresponding tunnelling probability. We have shown a rough estimate of the decay probability in Appendix I and found it negligibly small.

Before we find the predictions of the model, let us summarize the important conditions or assumptions which we need to ensure in order to realise successful inflation. They are: (i) at the onset of inflation and afterwards the energy of the universe is dominated by
 $\phi$ field which implies $\lambda_\chi\chi^4< \frac{1}{2} m^2\phi^2$, (ii) the $\chi$ field during inflation is massive compared to the Hubble scale {\it i.e.} $m_\chi^2>H_{\rm Inf}^2$
so that we can integrate out the $\chi$ field during inflation, (iii) we keep value of $\chi$ field sub-Planckian.

Let us proceed to find out the values of the inflationary observables: spectral index ($n_s$) and tensor to scalar ratio ($r$) using the modified potential $V_{\rm Inf}^{\rm eff}$ in equation (\ref{eq:InfPeff}). The analytic expressions for $n_s$ and $r$ in our set up are obtained as (in $M_P=1$ unit)
\begin{align}
&\epsilon=\frac{1}{2}\Big(\frac{V_{\rm Inf}^{ \prime}}{V_{\rm Inf}}\Big)^2=\frac{2}{\phi^2}\Big[\frac{1-2\alpha\phi^2}{1-\alpha\phi^2}\Big]^2,\\
&\eta=\Big(\frac{V_{\rm Inf}^{ \prime\prime}}{V_{\rm Inf}}\Big)=\frac{2}{\phi^2}\Big[\frac{1-6\alpha\phi^2}{1-\alpha\phi^2}\Big],
\end{align}
where $V^\prime=\frac{\partial V}{\partial\phi}$. The number of e-foldings can be determined using
\begin{align}
N_e=\int_{\phi_{\rm end}}^{\phi^*}\frac{\phi(1-\alpha\phi^2)}{2(1-2\alpha\phi^2)}d\phi,
\end{align}
where $\phi_{\rm end}$ and $\phi^{*}$ represent the field value at the end of inflation and the point of horizon exit respectively. The number of e-fold is connected to the inflationary parameters through the following relation \cite{Liddle:2003as,Martin:2010kz,Dodelson:2003vq,Kolb:1990vq},
\begin{align}
 N_e\simeq63.3+\frac{1}{4}{\rm ln}[\epsilon]+\frac{1}{4}{\rm ln}\left[\frac{V_{\rm Inf}}{\rho_{\rm end}}\right]+\frac{1}{12}{\ln}\left[\frac{T_{\rm rh}^4}{\rho_{\rm end}}\right],
 \label{eq:Efold}
\end{align}
where $V_{\rm Inf}$ indicates the energy scale of inflation. The energy density at the end of inflation and reheating temperature of the universe  are  denoted by $\rho_{\rm end}$ and $T_{\rm rh}$ respectively. Once these are known, $N_e$ can be easily computed.
Now the spectral index and tensor to scalar ratio in slow-roll inflation model are defined as
\begin{align}
&n_s=1-6\epsilon+2\eta,\label{nS}\\
&r=16\epsilon\label{TS}.
\end{align}
The curvature perturbation spectrum is given by
\begin{align}
P_S=\frac{V_{\rm Inf}}{24\pi^2\epsilon}=\frac{m^2\phi^4}{96\pi^2}\frac{(1-\alpha\phi^2)^3}{(1-2\alpha\phi^2)^2}.
\end{align}
The observed value of $P_S$ is found to be $2.2\times 10^{-9}$ at a pivot scale $k^* \sim 0.05$ Mpc $^{-1}$ \cite{Aghanim:2018eyx}.
Using equations (\ref{nS}-\ref{TS}) for different values of $\alpha$ we numerically estimate
 the magnitudes of $n_S$ and $r$ as shown in table \ref{tab:Predictions} for number of e-folds $N_e=60$. The choice of $N_e$ is made following equation (\ref{eq:Efold}) considering $V_{\rm Inf}^{1/4}\sim 10^{16}$ GeV, $\rho_{\rm end}^{1/4}\sim 10^{12}$ GeV, $\epsilon\sim 0.001$ and $T_{\rm rh}\sim 10^{14}$ GeV (the estimate of $T_{\rm rh}$ will be shown shortly in the current section).
The corresponding values of $m$ (in $M_P=1$ unit) are obtained in table \ref{tab:Predictions} using the observed value of $P_S$.
\begin{table}[h]
\centering
\begin{tabular}{ | l | l | l | l | l | }
\hline
No. of e-folds & m & $\alpha$ & $n_s$ & $r$ \\ 
 \hline
\multirow{3}{*}{$N_e=60$}
 &$5.94\times 10^{-6}$ & 0.0003 & 0.9661 & 0.1174 \\
  &$5.83\times 10^{-6}$ & 0.0007 & 0.9652 & 0.0970 \\
 &$5.59\times 10^{-6}$ & 0.0011 & 0.9620 & 0.0760 \\ \hline
\end{tabular}
\caption{Predictions for the modified version of chaotic inflation model in $M_P=1$ unit.}
\label{tab:Predictions}
\end{table}
We also show the predictions of our proposed model in figure \ref{fig:InfPo} (right panel) 
by varying $\alpha$ from 0.0001 to 0.0015. It can be
concluded from right panel of figure \ref{fig:InfPo} that with the increase of $\alpha$, value of $r$ can be reduced (in comparison with minimal form of chaotic inflation) to be consistent with Planck TT, TE, EE+low E + lensing results \cite{Akrami:2018odb}. However the model is ruled out if we consider Planck TT, TE, EE + low E + lensing +BICEP 2/Keck Array (BK15) + BAO data \cite{Akrami:2018odb} which is much stringent than earlier. {In addition, unboundness exists for the effective inflationary potential (equation (\ref{eq:InfPeff})) at large $\phi$ value which gives rise to metastability issue \cite{Ballesteros:2015iua}. Below we will see that addition of a mere higher dimensional term to the inflationary sector will alleviate this problem and in addition make the inflationary predictions consistent with Planck+BK15+BAO bounds as well.}

\subsection{Inflation with higher dimensional operator}
\begin{figure}[h]
 \centering
 \includegraphics[height=5cm,width=8cm]{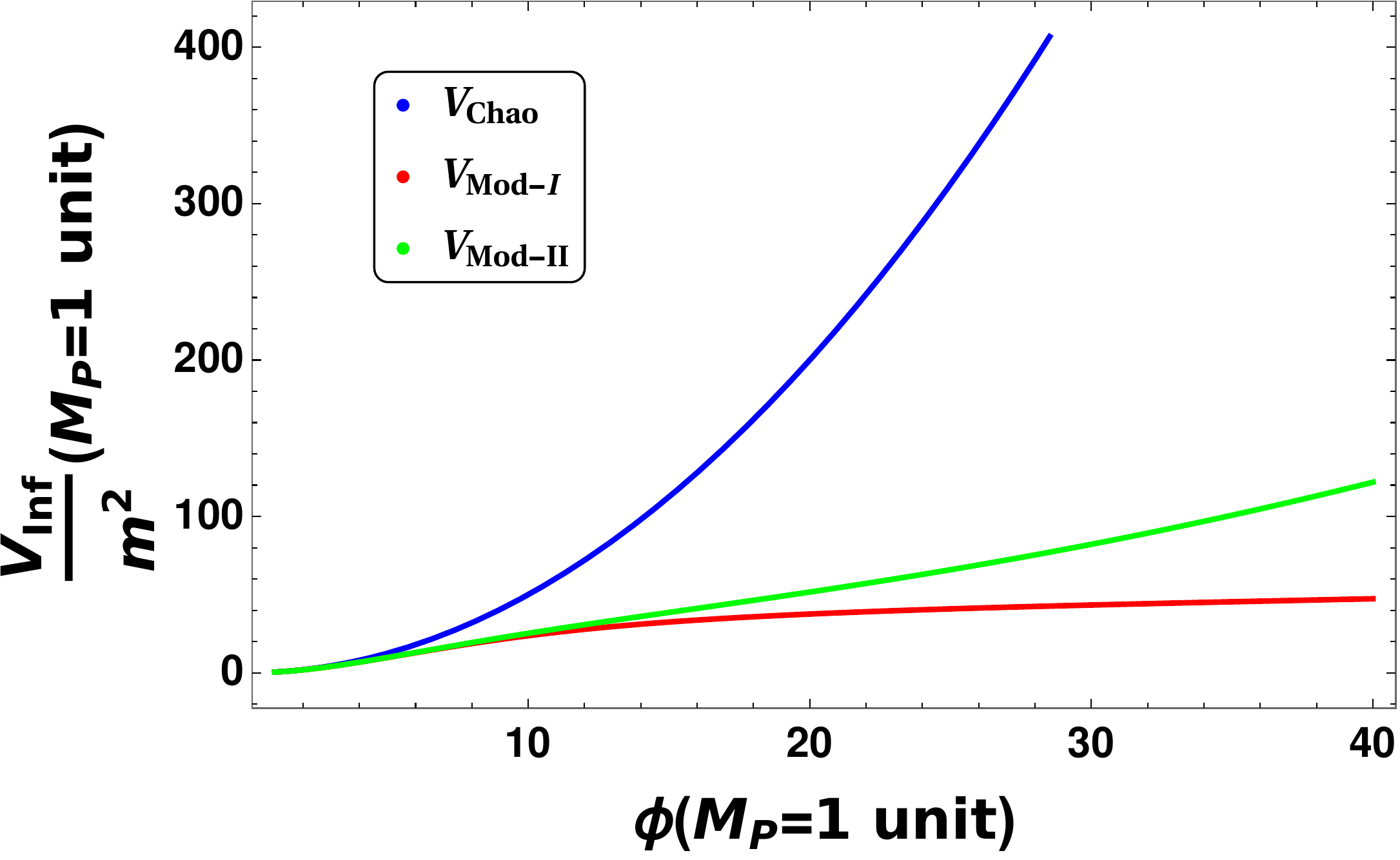}
 \caption{Sketch of $V_{\rm Inf}^{\rm eff}$ in equation (\ref{eq:rePo}) for reference points I and II of table \ref{tab:Predictions2} {\it i.e.} $\{m,\beta_1,\beta_2\}$ $\rightarrow$ $\{6.72\times 10^{-6}, 1.78\times 10^{-9},1.12\times 10^{-10}\}$ (green) and $\{7.65\times 10^{-6}, 1.78\times 10^{-9},1.25\times 10^{-10}\}$ (red). For comparison purpose, we also include the potential structure of the minimal chaotic potential (blue).}
 \label{fig:Po1}
\end{figure}

\begin{figure}[h]
 \centering
 \includegraphics[height=6cm,width=8cm]{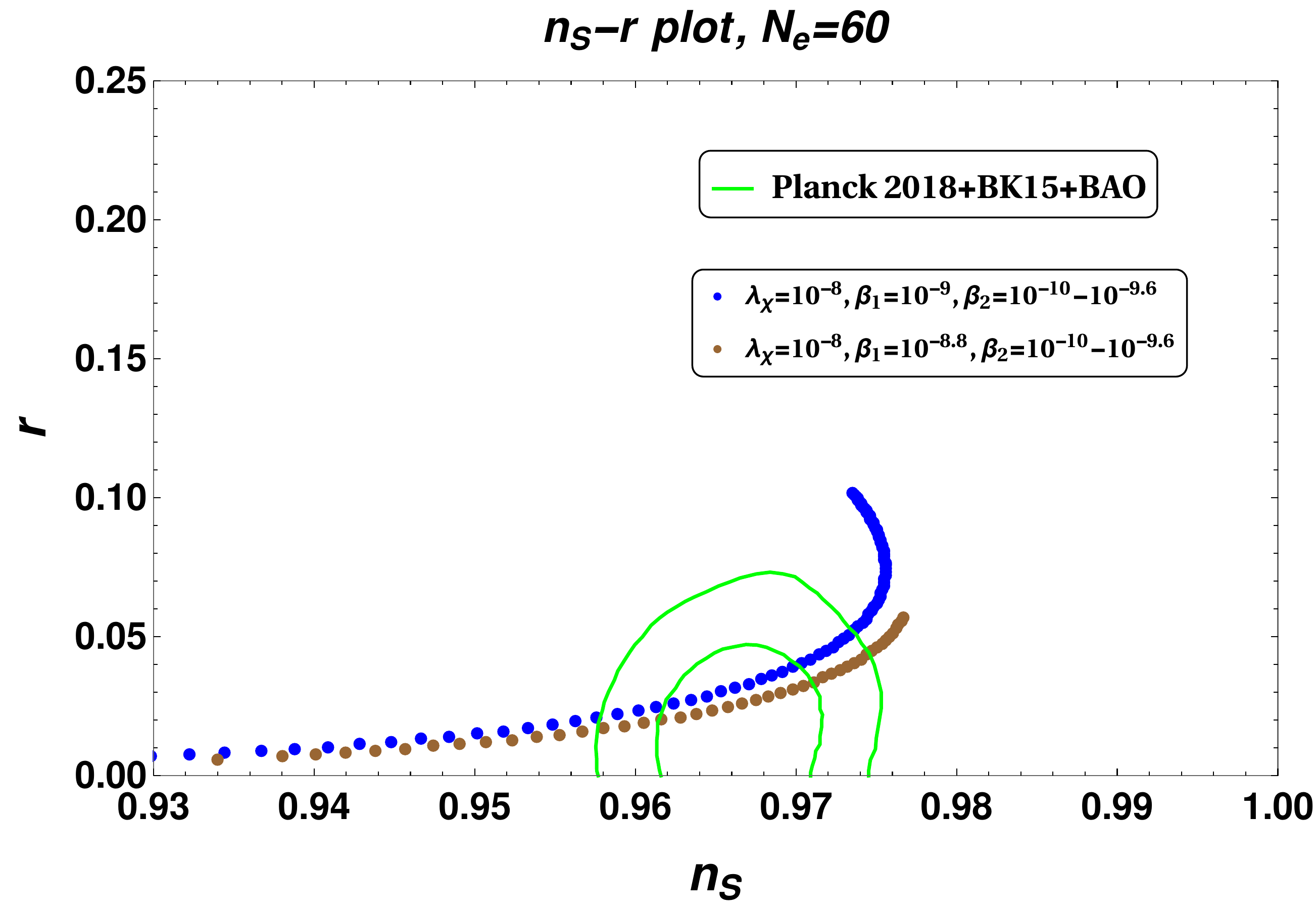}
  \includegraphics[height=6cm,width=8cm]{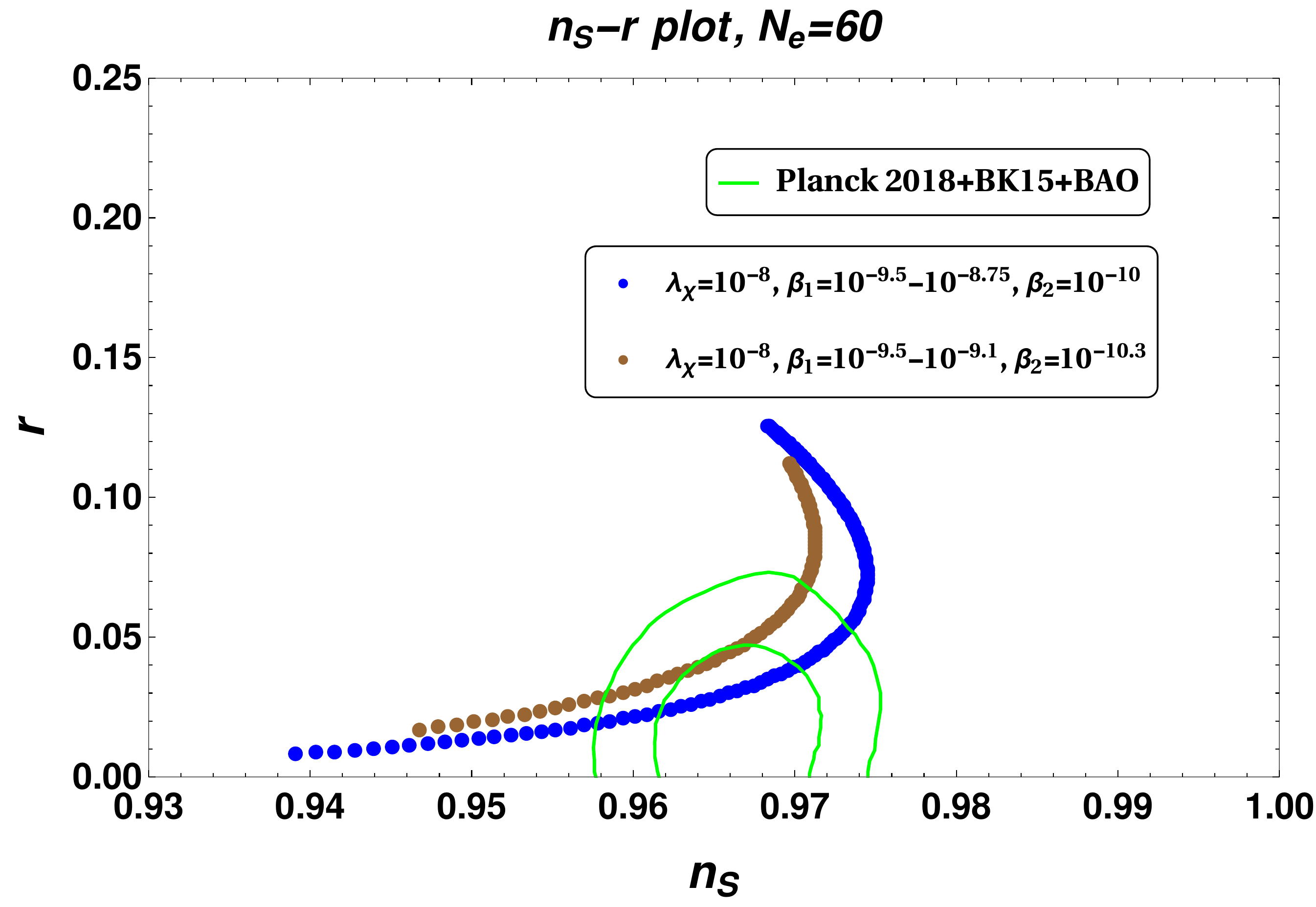}
 \caption{$n_s$-$r$ contour lines for $N_e=60$ as obtained from the redefined inflationary potential of equation \eqref{eq:rePo}. In left panel we fix $\beta_1$ and vary $\beta_2$ while the right panel shows the reverse case. The value of $\lambda_\chi$ is kept fixed at $10^{-8}$. We also include Planck TT, TE, EE+lowE+lensing+BK15+BAO $1\sigma$ and $2\sigma$ allowed contours.}
 \label{fig:predictions1}
\end{figure}

In an effort to make the model consistent with Planck+BK15+BAO data we incorporate a higher dimensional operator (which is perfectly allowed from the charge assignments in table \ref{tab:charge_assignments}) in the inflationary potential (equation (\ref{InfP})) given by,
\begin{align}
 V^{\rm Inf}_{\rm HO}=\frac{c_2}{8\Lambda^2}\Big(\chi^2-\frac{v_\chi^2}{2}\Big)^2\phi^2,
 \label{HO}
\end{align}
{where we assume $c_2$ to be real and positive.} We take $\Lambda=M_P$ as the natural cut off scale of the theory. The total inflationary scalar
potential is written as $V^{\rm Inf}_T=V(\phi,\chi)+V^{\rm Inf}_{\rm HO}$. Similar to the earlier case,
here also we integrate out the heavier $\chi$ field which again receives a negative  mass-squared larger than the $\mathcal{O}(H_{\rm Inf}^2)$ at the onset of inflation. Then we obtain the effective inflationary potential which is (in $M_P=1$ unit),
\begin{align}
 V_{\rm Inf}^{\rm eff}=\frac{m^2\phi^2}{2}\Big[1-\frac{\beta_1\phi^2}{16(\lambda_\chi+\beta_2\phi^2)}\Big],\label{eq:rePo}
\end{align}
where $\beta_1=\frac{2 c_1^2}{m^2}$ and $\beta_2=\frac{c_2}{2 M^2_P}= \frac{c_2}{2}$ and we also consider $v_\chi^2\ll \frac{c_1}{\lambda_\chi}$ (in $M_P=1$ unit).
Using this effective potential one can calculate the inflationary predictions $n_s$ and $r$ using equations (\eqref{nS}-\eqref{TS}) by varying $\beta_1$ and $\beta_2$ for a fixed value of $\lambda_\chi\sim 10^{-8}$. The parameter $m$ would be determined from the observed value of curvature perturbation spectrum ($P_S$). It is important to note that {the inflationary potential in equation (\ref{eq:rePo}) has no maxima at large values of $\phi$ provided $\beta_1<16\times\beta_2$ with $\beta_1$ and $\beta_2$ being real and positive. This inequality is obtained using the condition $V^{\rm eff}_{\rm Inf}>0$ for any arbitrary value of $\phi$. Considering the inequality $\beta_1<16\times \beta_2$ is satisfied, the effective inflationary potential (equation \eqref{eq:rePo}) due to the presence of higher dimensional operator $V_{\rm HO}^{\rm Inf}$ in equation (\ref{HO})
is monotonically increasing function of the $\phi$. Therefore, the inflaton $\phi$ can naturally roll towards the minimum $\phi=0$ from any arbitrary large value. This notable feature of the effective inflationary potential makes the set up more favored unlike the previous case ($\beta_2=0$) where the effective inflationary potential suffers from the issue of unboundeness at large $\phi$.

We have shown a sketch of the effective inflationary potential (equation \ref{eq:rePo}) in figure \ref{fig:Po1} for two benchmark points which exhibits that the effective inflationary potential (equation (\ref{eq:rePo})) is much flatter than the one in minimal chaotic model with only $m^2\phi^2$ potential. Here the flattening starts to take place near $\phi\sim 8 M_P$. We estimate $N_e$ to be 60 corresponding to the inflationary energy scale $\sim 10^{16}$ GeV and reheat temperature $\sim 10^{14}$ GeV similar to the earlier case. We also take into account the limit $\beta_1<16\times \beta_2$ while scanning the parameter space.} In figure \ref{fig:predictions1} we show the predictions of the proposed model in presence of the higher dimensional term (equation (\ref{HO})) in $n_s-r$ plane for $N_e=60$. We vary $\beta_1$ for a definite $\beta_2$ in right panel while the left panel shows the
effect of varying $\beta_2$ for a fixed $\beta_1$. It is clear that both $\beta_1$ and $\beta_2$ in equation (\ref{eq:rePo}) help in lowering the value of $r$ such that the inflationary predictions (both $n_s$ and $r$) successfully fall within the stringent Planck TT, TE+low E+lensing+BK15+BAO bounds. We tabulate two reference points for $N_e=60$ in table \ref{tab:Predictions2} which show the numerical estimates of the inflationary predictions following the redefined potential in equation \eqref{eq:rePo}.
\begin{table}[h]
\centering
\begin{tabular}{| l | l | l | l | l | l | }
\hline
No. of e-folds & $m$ & $\beta_1$ & $\beta_2$ & $n_s$ & $r$\\ \hline
\multirow{2}{*}{$N_e=60$}
 &$6.72\times 10^{-6}$ & $1.78\times 10^{-9}$ & $1.12\times 10^{-10}$ & 0.970 & 0.038\\
  &$7.65\times 10^{-6}$ & $1.78\times 10^{-9}$ & $1.25\times 10^{-10}$ & 0.975 & 0.057 \\ \hline
\end{tabular}
\caption{Predictions for the modified version of chaotic inflation model with higher order contribution in $M_P=1$ unit. We have fixed $\lambda_\chi=10^{-8}$.}
\label{tab:Predictions2}
\end{table}

 Now considering $c_1=\sqrt{\frac{\beta_1}{2}}m\sim 1.11\times 10^{-11}$ and
 $\lambda_\chi\sim 10^{-8}$ as in benchmark point I of table \ref{tab:Predictions2} and $\phi\sim15 M_P$ during inflation, we find $v_\chi\ll 0.5 M_P $  in order to maintain the 
validity of our approximation in equation (\ref{chiV}) and equation \eqref{eq:rePo}. In addition if we consider
mass of the extra scalar field ($m_\chi$) to be larger than $\mathcal{O}(10)$ TeV, a lower
bound on $v_\chi$ can also be obtained which is $v_\chi\gtrsim 10^8$ GeV for $\lambda_\chi\sim 10^{-8}$. The
 upper bound on $v_\chi$ can be further improved by taking into account of
the high scale validity of the proposed model. A stronger upper bound  $v_\chi\lesssim 10^{14}$ GeV can be found for $\lambda_\chi\sim 10^{-8}$ if we consider the high scale stability
of electroweak vacuum during and after inflation \cite{Saha:2016ozn}. Moreover, the high scale validity 
of the model also constrains the magnitude of $\lambda_{\chi H}$ which should be of same order as $\lambda_\chi$ so that
it does not alter the RG running of $\lambda_\chi$ to a great extent.  Another important point to note in the proposed set up is the order of magnitudes of $\lambda_{\phi H}$ and $\lambda_{\phi\eta}$, the coupling coefficients of inflaton with SM Higgs
and $\eta$ respectively in equation (\ref{eq:PotTot}). Following \cite{Lebedev:2012sy}, it can be shown that $\lambda_{\phi H},\lambda_{\phi\eta}>\mathcal{O}(10^{-6})$ can spoil the required flatness of the chaotic inflationary potential by inducing large radiative corrections. Therefore we keep both $\lambda_{\phi H}$ and $\lambda_{\phi\eta}$ smaller than $\mathcal{O}(10^{-6})$ in our scenario.

\subsection{Reheating:} 
Once inflation ends, $\phi$ field rolls down and oscillates about its minimum $\phi=0$. Then
the reheating process starts to take place. In our model we have various possibilities of energy transfer of inflaton to the relativistic degrees of freedom through $\phi^2 \lvert H \rvert^2,~ \phi^2\chi^2,~\phi^2 \lvert \eta \rvert^2$ interaction terms. If we consider $\lambda_{\phi H}\gg \lambda_{\phi \chi}, \lambda_{\phi\eta}$, production of $H$ fields at zero crossing
of $\phi$ field during its oscillation will occur dominantly. Now
$H$ field can decay to SM gauge bosons and fermions with coupling strength determined by the SM gauge coupling constant and the Yukawa couplings respectively.
Hence it is possible that the produced $H$ field from $\phi$ may decay to SM fields before the oscillating field $\phi$ returns back to the minimum of the
potential from its maximum value during oscillation. This event is known as instant preheating \cite{Felder:1999pv,Felder:1999pv1,ArmendarizPicon:2007iv,Desroche:2005yt,Allahverdi:2011aj}.
In the process, the effective mass of $H$ field grows as $m^2_H = \lambda_{\phi H}\phi^2$ when the field $\phi$ rolls up from the minimum of the effective potential. Then the effectively heavy Higgs field decays to SM fields at the moment when it has the greatest mass, i.e. when $\phi$ reaches its maximal value. 
The number density of produced $H$ fields at zero crossing of inflaton
can be obtained as \cite{Felder:1999pv,Felder:1999pv1}
\begin{align}
 n_H \simeq \frac{\lambda_{\phi H}^{3/4} \dot{\phi_0}^{3/2}}{8\pi^3},
\end{align}
where $\dot{\phi_0}$ is the velocity of the inflaton around $\phi=0$.
For chaotic inflation model with quadratic potential, the amplitude of first oscillation is $\sim 0.1 M_P$. Hence the effective mass of
produced Higgs field will be around $0.1 \sqrt{\lambda_{\phi H}} M_P$ which implies Higgs will be non relativistic at that moment. Then the total energy density of the Higgs field can be approximated as \cite{Felder:1999pv,Felder:1999pv}
\begin{align}
 \rho_H=m_H n_H \sim 10^{-14} \lambda_{\phi h}^{5/4} M_P^4.
\end{align}

Now the created $H$ fields will decay completely at the moment $\phi$ reaches its maximum during oscillation if the condition $(\Gamma_H)^{-1}\sim \Delta t$ is obeyed, where $\Delta t$ is the required time
for $\phi$  to reach its maximal value during the oscillation. Suppose the decay width of $H$ to SM particles is provided by
\begin{align}
 \Gamma_H=\frac{\delta^2 m_H}{8\pi},
\end{align}
where $\delta$ is the coupling of Higgs with SM particles.
In \cite{Felder:1999pv}, it is shown that for $\delta^2\lambda_{\phi H}^{1/2}\sim 5\times 10^{-4}$, indeed the decay products of $H$ can dominate the energy density of the universe. This can indeed happen if $\lambda_{\phi H}\gtrsim 10^{-8}$, which holds in our model.
It is to be noted that the energy dilution of the $\phi$ field can also happen through the  perturbative decay $\phi\rightarrow \chi \eta \eta$ as followed from equation (\ref{eq:PotTot}) with the corresponding decay rate 
\begin{align}
 \Gamma_{\phi\rightarrow \chi \eta \eta}\simeq\frac{\pi \lambda_R^2}{2 m}\Big[\frac{m}{8}\sqrt{m^2-4m_\chi^2}-\frac{m_\chi^2}{2}\textrm{log}\Big\{\frac{1}{2m_\chi}(m+\sqrt{m^2-4 m_\chi^2})\Big\}\Big].
\end{align}

Considering the instant preheating to be the dominant process to transfer the energy of inflaton to relativistic particles, we can try to
find a numerical estimate of the reheat temperature. After several oscillation of the inflaton, the complete transfer of the inflaton energy via $\phi\rightarrow H \rightarrow f \bar{f}$ occurs. Then we can use the relation 
$\rho_T=\frac{\pi^2}{30}g_* T^4$ where $g_*$
is the number of relativistic degrees of freedom in the thermal bath. At this moment we can assign a reheat temperature to the universe as
\begin{align}
 T_{\rm rh}=\left(\frac{30 }{\pi^2g_*}\right)^{1/4} \rho_0^{1/4}\simeq 10^{14} {\rm GeV}
\end{align}
where we take $\rho_0\simeq \frac{1}{2} m^2\phi_0^2$ with $\phi_0$ is the initial amplitude of $\phi$ oscillation.
\section{Dark Matter}
\label{sec3}
Here we present the detailed analysis of DM phenomenology in our model. As pointed out earlier, we consider the fermion singlet field $\psi$ as the DM candidate. From the Lagrangian of the model involving $\psi$ upto dimension five level, it is clear that $\psi$ does not have any decay mode and hence stable. This is also ensured by the $Z^{\prime}_4$ symmetry which remains unbroken even after $\chi$ field acquires non-zero VEV \footnote{It is expected that $\chi$ field having nonzero VEV will mix with the SM Higgs doublet. However, considering the largeness of $v_\chi\gtrsim 10^9$ GeV and small $\lambda_{\chi H}\sim \lambda_\chi\simeq 10^{-9}$, the mixing angle turns out to be too small any phenomenological consequences.}. Thus, the contribution to DM is expected to come entirely from $\psi$ as long as it is the lightest particle charged under the unbroken $Z^{\prime}_4$ symmetry. Since $\psi$ is a gauge singlet and it does not have any renormalisable interactions with other particles of the model, it is natural that its interactions with the visible sector particles will remain out of thermal equilibrium in the early universe, a requirement to realise the freeze-in DM scenario. Before the scalar fields acquire VEV, the production of DM can occur from $2 \rightarrow 2$ scattering processes. However, the effective vertex of such scattering diagrams are Planck scale suppressed. This leads to a small ultra-violet (UV) freeze-in contribution to DM abundance \cite{Elahi:2014fsa}. After the scalar field $\chi$ acquires non-zero VEV, there can be an effective two body decay contributions to DM production (infrared freeze in). This can arise from $\chi \rightarrow \psi \psi$ and $\eta \rightarrow \psi \psi$ with the effective coupling is governed by the ratio $v_{\chi}/M_P$ which can be as small as $10^{-10}$ for $v_{\chi} \approx 10^9$ GeV. Such couplings are quite generic in FIMP scenarios where DM is produced from such two body decays of mother particles \cite{Hall:2009bx}. Out of these two contributions, the process $\eta \rightarrow \psi \psi$ will dominant as $\chi$ has other decay modes whose couplings are not Planck scale suppressed. For example,  $\chi$ field can decay dominantly into a pair of Higgs or a pair of $\eta$. This tree level decays make the field $\chi$ to lose its abundance very quickly. On the other hand, the $\eta$ field gets produced in the thermal bath in the early universe and eventually can freeze-out from the thermal plasma at a later stage. This is similar to the superWIMP scenario \cite{Feng:2003uy} where a metastable WIMP decays into a super-weakly interacting dark matter at late epochs. This can happen because $\eta$ can decay only into $\psi$ and the corresponding coupling is very small, thereby allowing the freeze-out to occur earlier. Thus, $\eta$ can decay to $\psi$ while it is in equilibrium and also after its thermal freeze-out. 
The decay width of mother particle $\eta$,
\begin{align}
\Gamma_{\eta\rightarrow \bar{\psi} \nu_R}=\frac{{ y_{eff}}^2 \left(m{_\eta}^2-m_{\psi }^2\right) \left(1-\frac{m_{\psi }^2}{m{_\eta}^2}\right)}{8 \pi  m{_\eta}},
\label{DW:2}
\end{align}
where $y_{\rm e
ff}=\frac{\xi v_{\chi}}{M_{P}}$ is the effective coupling of $\eta \psi \nu_R$ vertex.

The relic abundance of both $\eta$ and $\psi$ can be found  solving the following set of Boltzmann equations which can be expressed as  
\begin{align}
&\frac{dY_{\eta}}{dx} = -\frac{4 \pi^2}{45}\frac{M_{P} m_{\rm \eta}}{1.66}\frac{\sqrt{g_{\star}(x)}}{x^2}\langle\sigma v\rangle_{\eta}^T
\left( Y_{\eta}^2-Y_{\eta}^{{eq}^2}\right) 
 - \frac{ M_{P}}{1.66} \frac{x\sqrt{g_{\star}(x)}}{m_{\rm \eta}^2\ g_s(x)} \Gamma_{\eta \rightarrow \bar{\psi} \nu_R}  \ Y_{\eta},
\label{BE:1}\\
&\frac{dY_\psi}{dx} =\frac{ M_{P}}{1.66} \frac{x\sqrt{g_{\star}(x)}}{m_{\eta}^2 \ g_s(x)} \Gamma_{\eta \rightarrow \bar{\psi} \nu_{R}}
\ Y_{\eta},
\label{BE:2}
\end{align}
where the comoving equilibrium number density of $\eta$ is given by
\begin{align}
 Y_{\eta}^{ eq}= 0.145\, \frac{g}{g_{s}(T)}\,\, \bigg(\frac{m_{\eta}}{T}\bigg)^{3/2}\,\, e^{-\frac{m_\eta}{T}},
\end{align}
with $T$ being the temperature of the thermal bath and g is the internal degrees of freedom of $\eta$.
The equation (\ref{BE:1}) corresponds to the evolution of the comoving number density ($Y=\frac{n}{s}$) of mother particle ($\eta$) as a function of $x=\left(\frac{m_\eta}{T}\right)$ where the first term on the right hand side shows the contribution from the thermal bath to the Y$_\eta$ whereas the second term stands for the dilution due to the decay of $\eta$ to the DM particles. Similarly, the equation \eqref{BE:2} represents the evolution of the DM particles in the universe from the non-thermal contribution coming from the decay of $\eta$. The effective relativistic degrees of freedom  during thermal equilibrium and entropy degrees of freedom are denoted by the usual notations $g_*$ and $g_s$ respectively in equations (\ref{BE:1}) and (\ref{BE:2}). The $\langle\sigma v\rangle_\eta^T$ in equation (\ref{BE:1}) stands for the thermally averaged total annihilation cross section of $\eta$ field as provided in equation (\ref{eq:TotSigV}) of Appendix II. To estimate the relic abundance of DM, we will numerically solve the coupled differential equations (\ref{BE:1}) and (\ref{BE:2}) and use the following expression,
\begin{align}
 \Omega_{\rm DM} h^2=\frac{h^2\, { m_{DM}}\, s_0}{\rho_{crit}}Y_\psi^{(x=\infty)}=2.755\times 10^8\times \Bigg(\frac{m_{\rm DM}}{\rm GeV}\Bigg)\, Y_\psi^{(x=\infty)}.
 \label{omega:Y}
\end{align}
where $s_0$, $Y_\psi^{(x=\infty)}$ are the present entropy density and comoving number density respectively and $\rho_{crit}$ is the critical energy density of the universe. Also, here we identify $m_{\psi}$ as $m_{\rm DM}$. The parameter $h$ is defined as: $h$ = (Hubble Parameter)/(100 km ${\rm sec^{-1}\,\, Mpc^{-1}}$).

\begin{figure}[h]
 \centering
 \includegraphics[width=8cm]{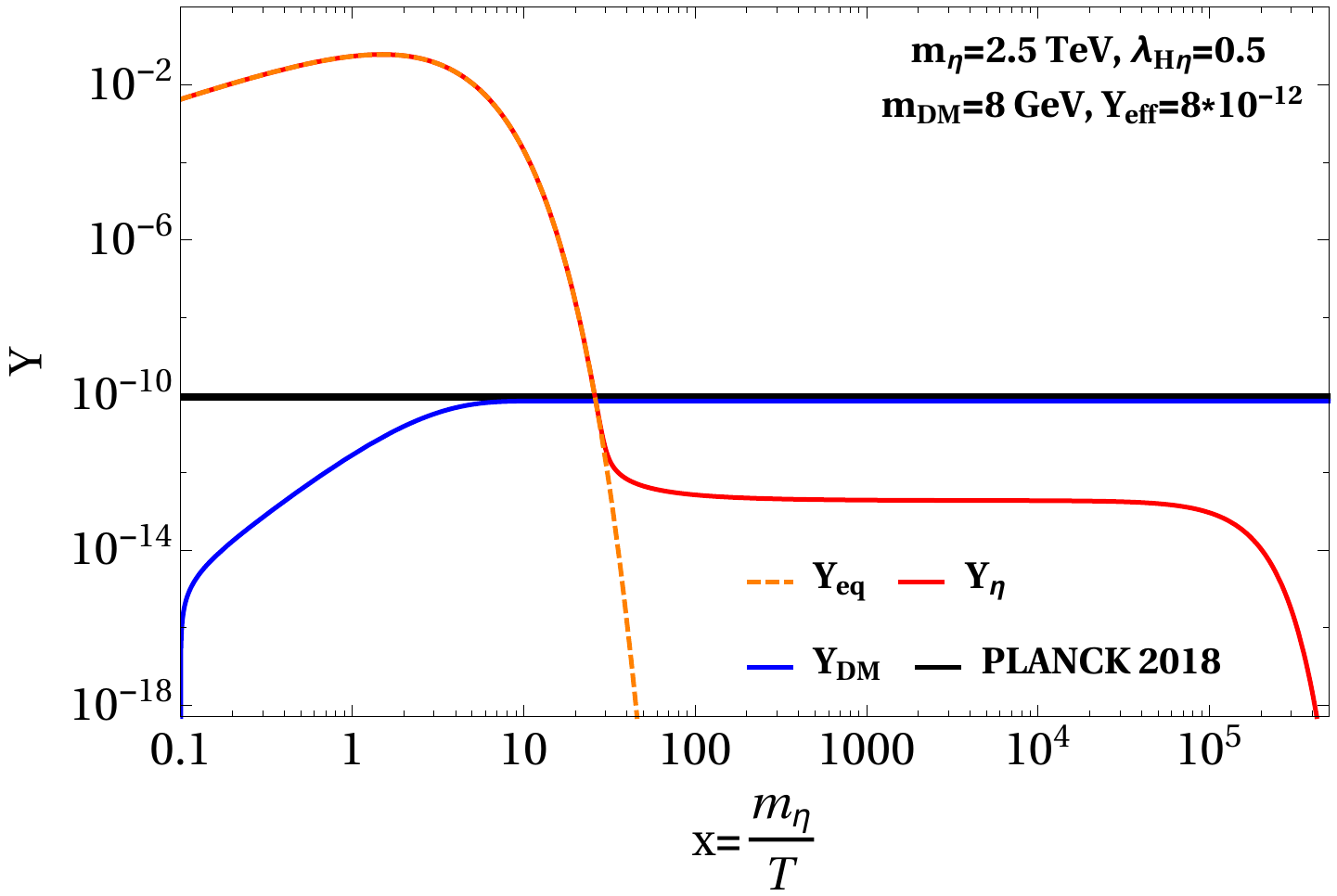}
 \caption{Variation of comoving number density $Y_{\rm DM}$ (same as $Y_\psi$) as a function of x (= $\frac{m_{\eta}}{T}$) for benchmark values of the other parameters. The black solid line stands for the required $Y$ to produce experimentally observed relic abundance of DM \cite{Aghanim:2018eyx} for the chosen set of parameters.}
 \label{fig:Yvar}
\end{figure}

\begin{figure}[h]
 \centering
 \includegraphics[width=8cm]{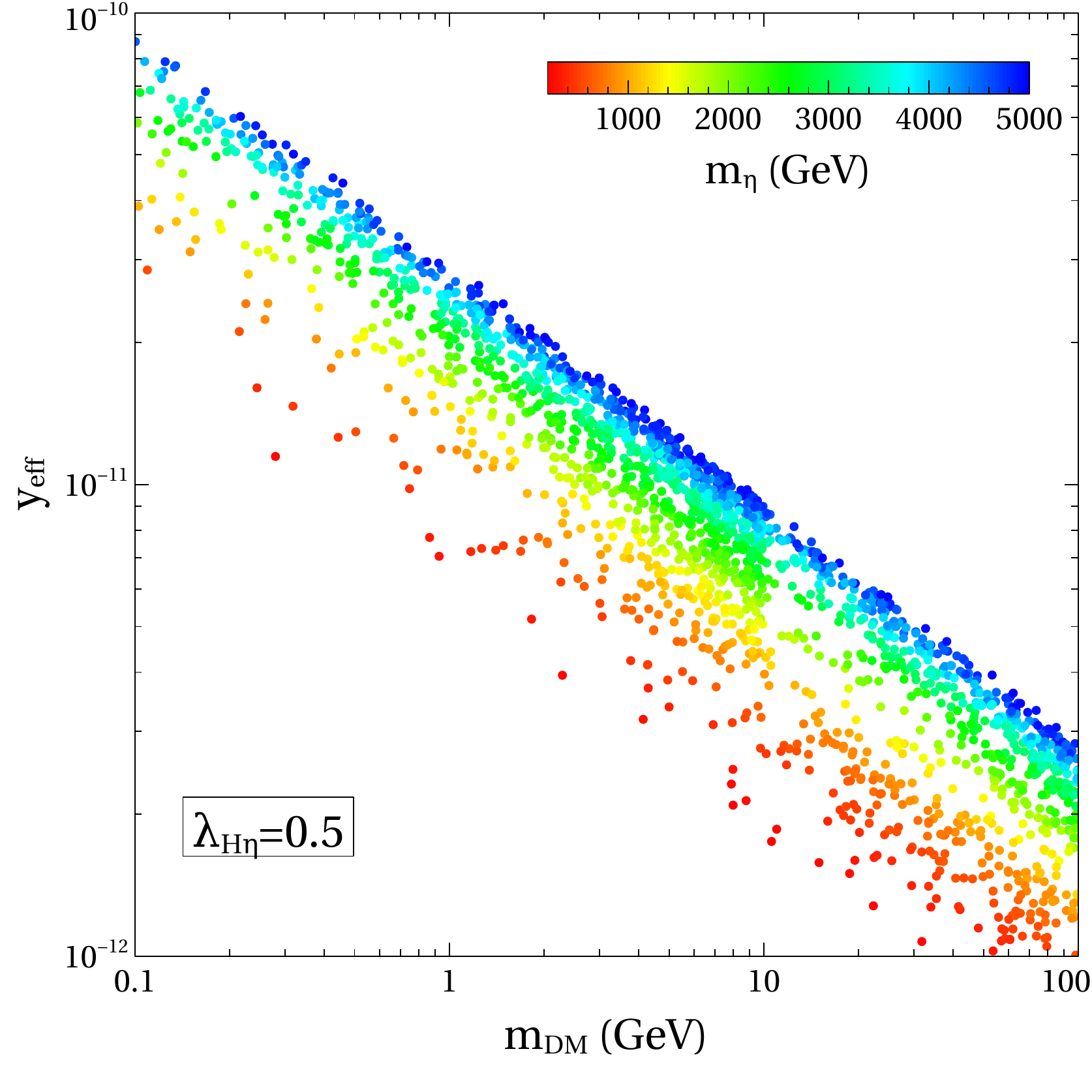}
 \caption{Allowed parameter space in $m_{\rm DM}-y_{\rm eff}$ plane where the variation of $m_{\eta}$ is shown through colour coding.}
 \label{fig:para}
\end{figure}

Before performing the numerical analysis we note down the parameters
which serve important roles in determining the total relic abundance of DM:
\begin{align}
\{m_{\rm DM},~m_\eta,~y_{\rm eff}, \lambda_{ H\eta}\}\label{eq:para}.
\end{align}
In figure \ref{fig:Yvar} we have shown the variation of the comoving number densities 
for both the fields $\eta$ and $\psi$ as a function of $x$ (=$\frac{m_{\eta}}{T}$) for a specific choice of values of the relevant
 parameters as mentioned in the figure. The evolution of Y$_\eta$ (red line) clearly shows that $\eta$ was in thermal bath in the early universe and went out of equilibrium through usual freeze-out mechanism and at some later stage continues to decay to DM particle $\psi$. The blue line in figure \ref{fig:Yvar} exhibits the evolution of $Y_{\rm \psi}$ which shows that initially the $\psi$ abundance was very small and increases gradually from the decay of the mother particle $\eta$. 
Figure \ref{fig:para} shows the relic density allowed parameter space in $m_{\rm DM}$-$y_{\rm eff}$ plane and the coloured bar represents the variation of m$_{\eta}$. Here we have varied the DM mass from 100 MeV to 100 GeV, $m_{\eta}$ from 500 GeV to 5 TeV and the $y_{\rm eff}$ from 10$^{-12}$ to 10$^{-8}$. It can be viewed that for smaller values of $m_{\rm DM}$ we need larger $y_{\rm eff}$ considering a fixed value of $m_\eta$.
 As mass of DM increases, the required number density of DM {\it i.e.} $Y_\psi$ (proportional to $\Gamma_\eta$) has to be smaller to satisfy the correct relic abundance (see equation \eqref{omega:Y}), and hence the decrease in $y_{\rm eff}$ is observed from figure \ref{fig:para}. Also, it is seen from figure \ref{fig:para} that the increase in $m_\eta$ raises $y_{\rm eff}$ in order to have correct amount of relic density. This can be understood by looking at equation (\ref{BE:2}), where larger $m_\eta$ causes suppression in the value of $Y_{\rm \psi}$ and hence it requires comparatively large $y_{\rm eff}$.

\section{Conclusion}
\label{sec4}
We have proposed a unified framework to account for non-thermal dark matter, modified chaotic inflation and sub-eV Dirac neutrino mass by minimally extending the standard model. The minimal chaotic inflation scenario, being ruled out by Planck data, is modified by an additional scalar field to bring the predictions for inflationary parameters within allowed range. The same additional scalar field also assists in generating required tiny couplings for non-thermal DM as well as Dirac neutrino mass by virtue of Planck scale suppressed dimension five operators. We find that for suitable VEV of the additional assisting field of the order $10^9$ GeV, it is natural to generate Dirac neutrino Yukawa of the order $10^{-12}-10^{-10}$ which can then generate sub eV Dirac neutrino mass by virtue of neutrino coupling to the SM Higgs. Similar couplings generated for DM coupling to its mother particle also makes the realisation of FIMP dark matter scenario natural. After showing the validity of modified chaotic inflation scenario for suitable benchmark choices of parameters, we numerically find the parameter space that can generate the correct FIMP DM abundance by scanning over DM mass, mother particle mass as well as their couplings. Future cosmology data should be able to make this model go through further scrutiny, specially in terms of the inflationary observables. Also the model can be falsified by observation of neutrinoless double beta decay which will rule out the pure Dirac nature of light neutrinos, as proposed in this model.

\acknowledgments
DB acknowledges the support from Indian Institute of Technology Guwahati start-up grant (reference number: xPHYSUGI-ITG01152xxDB001), Early Career Research Award from Science and Engineering Research Board (SERB), Department of Science and Technology (DST), Government of India (reference number: ECR/2017/001873) and Associateship Programme of IUCAA, Pune. DN and AKS thank Amit Dutta Banik for some useful discussions while carrying out the work.

\section*{Appendix I: Decay probability of false vacuum}
\appendix\label{appen1}
The decay probability of the false vacuum per unit time and unit volume can be calculated by adopting semi-classical method popularly known as bounce solution. A simpler approximate analytic form of the decay probability is given by \cite{Coleman:1977py,Sher:1988mj}
\begin{align}
 P_v\sim \phi_I^4e^{-S_4},
\end{align}
where $\phi_I$ is the starting field value of bounce and $S_4$ represents the Euclidean action for the bounce configuration. It can be shown that for a $\lambda\phi^4$ type potential $S_4$ approximately turns out to be \cite{Lee:1985uv,Arnold:1989cb}
\begin{align}
 S_4\simeq-\frac{8 \pi^2}{3\lambda}.
\end{align}
Now the volume of the past lightcone is estimated to be $\sim \left(\frac{e^{140}}{M_P}\right)^4$. With this, the total probability for nucleation of a bubble in present Hubble volume is \cite{Coleman:1977py,Sher:1988mj}
\begin{align}
 P_T=P_v\times \left(\frac{e^{140}}{M_P}\right)^4
\end{align}
In our case inflationary potential is dominated by $\phi^4$ term where the unboundeness starts to appear. Therefore it is legitimate to apply this simple method to calculate the tunnelling probability.
Considering $\phi_I\sim 30 ~M_P$ where the unboundness of the inflaton potential appears, and the quartic coupling coefficient of $\phi$ field as $\lambda\sim- m^2\alpha\simeq- 10^{-14}$, $P_T$ comes out to be much smaller than unity.

\newpage
\section*{Appendix II: annihilation cross sections of $\eta$}
Below we provide analytic expression of thermally averaged cross sections for all the possible annihilation processes of $\eta$ field.
\begin{align}
&\sigma_{\eta\eta\rightarrow f\overline{f}}=\frac{\lambda_{H\eta}^2 m_f^2 \left\{2 \left(s-2 m_f^2\right)-4 m_f^2\right\} \sqrt{\frac{s-4 m_f^2}{s-4 m_\eta^2}}}{16 \pi s \left\{\Gamma_h^2 m_h^2+\left(s-m_h^2\right)^2\right\}}, \\
&\sigma_{\eta\eta\rightarrow hh} = \frac{1}{16 \pi s}\sqrt{\frac{s-4 m_h^2}{s-4 m_\eta^2}} \Biggr[\frac{9 g^2 \lambda_{H\eta}^2 m_h^4 v^2}{4 m_W^2 \left\{\Gamma_h^2 m_h^2+\left(s-m_h^2\right)^2\right\}}+ \frac{3 g \lambda_{H\eta}^2 m_h^2 v}{m_w \sqrt{\Gamma_h^2 m_h^2 + \left(s-m_h^2\right)^2}}+4 \lambda_{H\eta}^2\Biggr],\\
&\sigma_{\eta\eta\rightarrow W^+ W^-}=\frac{g^2 \lambda_{H\eta}^2 m_W^2 v^2 \left\{\frac{\left(s-2 m_W^2\right)^2}{4 m_W^4}+2\right\} \sqrt{\frac{s-4 m_W^2}{s-4 m_{\eta}^2}}}{16 \pi s \left\{\Gamma_h^2 m_h^2+\left(s-m_h^2\right)^2\right\}},\\ 
&\sigma_{\eta \eta \rightarrow Z Z} = \frac{1}{16\pi s}\frac{g^2 \lambda_{H\eta}^2 m_Z^2 v^2 \left(\frac{\left\{s-2 m_Z^2\right)^2}{4 m_Z^4}+2\right\} \sqrt{\frac{s-4 m_z^2}{s-4 m_{\eta}^2}}}{\cos^2\theta_W \left\{\Gamma_h^2 m_h^2+\left(s-m_h^2\right)^2\right\}},
\end{align}
where we define,\\
$m_f$: mass of SM fermions,~~~~$m_W$: mass of $W$ boson,~~~
$m_Z$: mass of $Z$ boson,\\
$m_{\eta}$: mass of $\eta$ field,~~~~$m_{\rm DM}$: mass of DM,~~~~$g$: $SU(2)_L$ gauge coupling,\\$v$: vacuum expectation value of the SM Higgs,~~~~$\theta_W$: Weinberg angle,\\
$\Gamma_h$: decay width of SM Higgs, ~~~~$s$: center of mass energy.

 With all these inputs, the total annihilation cross section $\sigma_{T}$ and the thermal average of the cross-section $\langle\sigma v\rangle_\eta^T$ for $\eta$ can be written as
\begin{align}\label{eq:TotSigV}
 &\sigma_{T}=\sigma_{\eta\eta\rightarrow f\overline{f}}+\sigma_{\eta\eta\rightarrow hh}+\sigma_{\eta\eta\rightarrow W^+W^-}+\sigma_{\eta\eta\rightarrow ZZ}\\
&\langle \sigma v\rangle^T_\eta = \frac{1}{8\,m_\eta^4 \,T \, K^2_2(\frac{m_\eta}{T})}\int_{4m_\eta^2}^{\infty}\sigma_{T}\,(s-4m_{\eta}^2)\,\sqrt{s}\,K_1\bigg(\frac{\sqrt{s}}{T}\bigg)\,ds.
\end{align}


\begin{thebibliography}{115}
\expandafter\ifx\csname natexlab\endcsname\relax\def\natexlab#1{#1}\fi
\expandafter\ifx\csname bibnamefont\endcsname\relax
  \def\bibnamefont#1{#1}\fi
\expandafter\ifx\csname bibfnamefont\endcsname\relax
  \def\bibfnamefont#1{#1}\fi
\expandafter\ifx\csname citenamefont\endcsname\relax
  \def\citenamefont#1{#1}\fi
\expandafter\ifx\csname url\endcsname\relax
  \def\url#1{\texttt{#1}}\fi
\expandafter\ifx\csname urlprefix\endcsname\relax\def\urlprefix{URL }\fi
\providecommand{\bibinfo}[2]{#2}
\providecommand{\eprint}[2][]{\url{#2}}

\bibitem[{\citenamefont{Tanabashi et~al.}(2018)}]{Tanabashi:2018oca}
\bibinfo{author}{\bibfnamefont{M.}~\bibnamefont{Tanabashi}}
  \bibnamefont{et~al.} (\bibinfo{collaboration}{Particle Data Group}),
  \bibinfo{journal}{Phys. Rev.} \textbf{\bibinfo{volume}{D98}},
  \bibinfo{pages}{030001} (\bibinfo{year}{2018}).

  
  
\bibitem[{\citenamefont{Aghanim et~al.}(2018)}]{Aghanim:2018eyx}
\bibinfo{author}{\bibfnamefont{N.}~\bibnamefont{Aghanim}} \bibnamefont{et~al.}
  (\bibinfo{collaboration}{Planck}) (\bibinfo{year}{2018}),
  \eprint{1807.06209}.

\bibitem[{\citenamefont{Akerib et~al.}(2016)}]{Akerib:2015rjg}
\bibinfo{author}{\bibfnamefont{D.~S.} \bibnamefont{Akerib}}one of the ea
  \bibnamefont{et~al.} (\bibinfo{collaboration}{LUX}), \bibinfo{journal}{Phys.
  Rev. Lett.} \textbf{\bibinfo{volume}{116}}, \bibinfo{pages}{161301}
  (\bibinfo{year}{2016}), \eprint{1512.03506}.

\bibitem[{\citenamefont{Aprile et~al.}(2017)}]{Aprile:2017iyp}
\bibinfo{author}{\bibfnamefont{E.}~\bibnamefont{Aprile}} \bibnamefont{et~al.}
  (\bibinfo{collaboration}{XENON}), \bibinfo{journal}{Phys. Rev. Lett.}
  \textbf{\bibinfo{volume}{119}}, \bibinfo{pages}{181301}
  (\bibinfo{year}{2017}), \eprint{1705.06655}.

\bibitem[{\citenamefont{Cui et~al.}(2017)}]{Cui:2017nnn}
\bibinfo{author}{\bibfnamefont{X.}~\bibnamefont{Cui}} \bibnamefont{et~al.}
  (\bibinfo{collaboration}{PandaX-II}), \bibinfo{journal}{Phys. Rev. Lett.}
  \textbf{\bibinfo{volume}{119}}, \bibinfo{pages}{181302}
  (\bibinfo{year}{2017}), \eprint{1708.06917}.

\bibitem[{\citenamefont{Tan et~al.}(2016)}]{Tan:2016zwf}
\bibinfo{author}{\bibfnamefont{A.}~\bibnamefont{Tan}} \bibnamefont{et~al.}
  (\bibinfo{collaboration}{PandaX-II}), \bibinfo{journal}{Phys. Rev. Lett.}
  \textbf{\bibinfo{volume}{117}}, \bibinfo{pages}{121303}
  (\bibinfo{year}{2016}), \eprint{1607.07400}.

\bibitem[{\citenamefont{Hall et~al.}(2010)\citenamefont{Hall, Jedamzik,
  March-Russell, and West}}]{Hall:2009bx}
\bibinfo{author}{\bibfnamefont{L.~J.} \bibnamefont{Hall}},
  \bibinfo{author}{\bibfnamefont{K.}~\bibnamefont{Jedamzik}},
  \bibinfo{author}{\bibfnamefont{J.}~\bibnamefont{March-Russell}},
  \bibnamefont{and} \bibinfo{author}{\bibfnamefont{S.~M.} \bibnamefont{West}},
  \bibinfo{journal}{JHEP} \textbf{\bibinfo{volume}{03}}, \bibinfo{pages}{080}
  (\bibinfo{year}{2010}), \eprint{0911.1120}.

\bibitem[{\citenamefont{Bernal et~al.}(2017)\citenamefont{Bernal, Heikinheimo,
  Tenkanen, Tuominen, and Vaskonen}}]{Bernal:2017kxu}
\bibinfo{author}{\bibfnamefont{N.}~\bibnamefont{Bernal}},
  \bibinfo{author}{\bibfnamefont{M.}~\bibnamefont{Heikinheimo}},
  \bibinfo{author}{\bibfnamefont{T.}~\bibnamefont{Tenkanen}},
  \bibinfo{author}{\bibfnamefont{K.}~\bibnamefont{Tuominen}}, \bibnamefont{and}
  \bibinfo{author}{\bibfnamefont{V.}~\bibnamefont{Vaskonen}},
  \bibinfo{journal}{Int. J. Mod. Phys.} \textbf{\bibinfo{volume}{A32}},
  \bibinfo{pages}{1730023} (\bibinfo{year}{2017}), \eprint{1706.07442}.

\bibitem[{\citenamefont{Guth}(1981)}]{Guth:1980zm}
\bibinfo{author}{\bibfnamefont{A.~H.} \bibnamefont{Guth}},
  \bibinfo{journal}{Phys. Rev.} \textbf{\bibinfo{volume}{D23}},
  \bibinfo{pages}{347} (\bibinfo{year}{1981}), \bibinfo{note}{[Adv. Ser.
  Astrophys. Cosmol.3,139(1987)]}.

\bibitem[{\citenamefont{Starobinsky}(1980)}]{Starobinsky:1980te}
\bibinfo{author}{\bibfnamefont{A.~A.} \bibnamefont{Starobinsky}},
  \bibinfo{journal}{Phys. Lett.} \textbf{\bibinfo{volume}{B91}},
  \bibinfo{pages}{99} (\bibinfo{year}{1980}), \bibinfo{note}{[,771(1980)]}.

\bibitem[{\citenamefont{Linde}(1982)}]{Linde:1981mu}
\bibinfo{author}{\bibfnamefont{A.~D.} \bibnamefont{Linde}},
  \bibinfo{journal}{Phys. Lett.} \textbf{\bibinfo{volume}{108B}},
  \bibinfo{pages}{389} (\bibinfo{year}{1982}).

\bibitem[{\citenamefont{Komatsu et~al.}(2011)}]{Komatsu:2010fb}
\bibinfo{author}{\bibfnamefont{E.}~\bibnamefont{Komatsu}} \bibnamefont{et~al.}
  (\bibinfo{collaboration}{WMAP}), \bibinfo{journal}{Astrophys. J. Suppl.}
  \textbf{\bibinfo{volume}{192}}, \bibinfo{pages}{18} (\bibinfo{year}{2011}),
  \eprint{1001.4538}.
  
  
\bibitem[{\citenamefont{Akrami et~al.}(2018)}]{Akrami:2018odb}
\bibinfo{author}{\bibfnamefont{Y.}~\bibnamefont{Akrami}} \bibnamefont{et~al.}
  (\bibinfo{collaboration}{Planck}) (\bibinfo{year}{2018}),
  \eprint{1807.06211}.

\bibitem[{\citenamefont{Linde}(1983)}]{Linde:1983gd}
\bibinfo{author}{\bibfnamefont{A.~D.} \bibnamefont{Linde}},
  \bibinfo{journal}{Phys. Lett.} \textbf{\bibinfo{volume}{129B}},
  \bibinfo{pages}{177} (\bibinfo{year}{1983}).

\bibitem{Martin:2013tda} 
  J.~Martin, C.~Ringeval and V.~Vennin,
  Phys.\ Dark Univ.\  {\bf 5-6}, 75 (2014)
  [arXiv:1303.3787 [astro-ph.CO]].
  
\bibitem[{\citenamefont{Senoguz and Shafi}(2008)}]{NeferSenoguz:2008nn}
\bibinfo{author}{\bibfnamefont{V.~N.} \bibnamefont{Senoguz}} \bibnamefont{and}
  \bibinfo{author}{\bibfnamefont{Q.}~\bibnamefont{Shafi}},
  \bibinfo{journal}{Phys. Lett.} \textbf{\bibinfo{volume}{B668}},
  \bibinfo{pages}{6} (\bibinfo{year}{2008}), \eprint{0806.2798}.

\bibitem[{\citenamefont{Enqvist and Karciauskas}(2014)}]{Enqvist:2013eua}
\bibinfo{author}{\bibfnamefont{K.}~\bibnamefont{Enqvist}} \bibnamefont{and}
  \bibinfo{author}{\bibfnamefont{M.}~\bibnamefont{Karciauskas}},
  \bibinfo{journal}{JCAP} \textbf{\bibinfo{volume}{1402}}, \bibinfo{pages}{034}
  (\bibinfo{year}{2014}), \eprint{1312.5944}.

\bibitem[{\citenamefont{Ballesteros and Tamarit}(2016)}]{Ballesteros:2015noa}
\bibinfo{author}{\bibfnamefont{G.}~\bibnamefont{Ballesteros}} \bibnamefont{and}
  \bibinfo{author}{\bibfnamefont{C.}~\bibnamefont{Tamarit}},
  \bibinfo{journal}{JHEP} \textbf{\bibinfo{volume}{02}}, \bibinfo{pages}{153}
  (\bibinfo{year}{2016}), \eprint{1510.05669}.

\bibitem[{\citenamefont{Pallis and Shafi}(2015)}]{Pallis:2014cda}
\bibinfo{author}{\bibfnamefont{C.}~\bibnamefont{Pallis}} \bibnamefont{and}
  \bibinfo{author}{\bibfnamefont{Q.}~\bibnamefont{Shafi}},
  \bibinfo{journal}{JCAP} \textbf{\bibinfo{volume}{1503}}, \bibinfo{pages}{023}
  (\bibinfo{year}{2015}), \eprint{1412.3757}.

\bibitem[{\citenamefont{Kaewkhao and Gumjudpai}(2018)}]{Gumjudpai:2016ioy}
\bibinfo{author}{\bibfnamefont{N.}~\bibnamefont{Kaewkhao}} \bibnamefont{and}
  \bibinfo{author}{\bibfnamefont{B.}~\bibnamefont{Gumjudpai}},
  \bibinfo{journal}{Phys. Dark Univ.} \textbf{\bibinfo{volume}{20}},
  \bibinfo{pages}{20} (\bibinfo{year}{2018}), \eprint{1608.04014}.
  
\bibitem{Tenkanen:2017jih} 
  T.~Tenkanen,
  JCAP {\bf 1712}, 001 (2017)
  [arXiv:1710.02758 [astro-ph.CO]].

\bibitem[{\citenamefont{Kasuya and Taira}(2018)}]{Kasuya:2018cxo}
\bibinfo{author}{\bibfnamefont{S.}~\bibnamefont{Kasuya}} \bibnamefont{and}
  \bibinfo{author}{\bibfnamefont{M.}~\bibnamefont{Taira}},
  \bibinfo{journal}{Phys. Rev.} \textbf{\bibinfo{volume}{D98}},
  \bibinfo{pages}{123515} (\bibinfo{year}{2018}), \eprint{1803.10571}.

\bibitem[{\citenamefont{Van~Dong et~al.}(2019)\citenamefont{Van~Dong, Huong,
  Camargo, Queiroz, and Valle}}]{Dong:2018aak}
\bibinfo{author}{\bibfnamefont{P.}~\bibnamefont{Van~Dong}},
  \bibinfo{author}{\bibfnamefont{D.~T.} \bibnamefont{Huong}},
  \bibinfo{author}{\bibfnamefont{D.~A.} \bibnamefont{Camargo}},
  \bibinfo{author}{\bibfnamefont{F.~S.} \bibnamefont{Queiroz}},
  \bibnamefont{and} \bibinfo{author}{\bibfnamefont{J.~W.~F.}
  \bibnamefont{Valle}}, \bibinfo{journal}{Phys. Rev.}
  \textbf{\bibinfo{volume}{D99}}, \bibinfo{pages}{055040}
  (\bibinfo{year}{2019}), \eprint{1805.08251}.

\bibitem[{\citenamefont{Harigaya et~al.}(2016)\citenamefont{Harigaya, Ibe,
  Kawasaki, and Yanagida}}]{Harigaya:2015pea}
\bibinfo{author}{\bibfnamefont{K.}~\bibnamefont{Harigaya}},
  \bibinfo{author}{\bibfnamefont{M.}~\bibnamefont{Ibe}},
  \bibinfo{author}{\bibfnamefont{M.}~\bibnamefont{Kawasaki}}, \bibnamefont{and}
  \bibinfo{author}{\bibfnamefont{T.~T.} \bibnamefont{Yanagida}},
  \bibinfo{journal}{Phys. Lett.} \textbf{\bibinfo{volume}{B756}},
  \bibinfo{pages}{113} (\bibinfo{year}{2016}), \eprint{1506.05250}.

\bibitem[{\citenamefont{Saha and Sil}(2017)}]{Saha:2016ozn}
\bibinfo{author}{\bibfnamefont{A.~K.} \bibnamefont{Saha}} \bibnamefont{and}
  \bibinfo{author}{\bibfnamefont{A.}~\bibnamefont{Sil}},
  \bibinfo{journal}{Phys. Lett.} \textbf{\bibinfo{volume}{B765}},
  \bibinfo{pages}{244} (\bibinfo{year}{2017}), \eprint{1608.04919}.

  \bibitem{Felder:1999pv} 
  G.~N.~Felder, L.~Kofman and A.~D.~Linde,
  Phys.\ Rev.\ D {\bf 59}, 123523 (1999)
  [hep-ph/9812289].
  
\bibitem[{\citenamefont{Choubey and Kumar}(2017)}]{Choubey:2017hsq}
\bibinfo{author}{\bibfnamefont{S.}~\bibnamefont{Choubey}} \bibnamefont{and}
  \bibinfo{author}{\bibfnamefont{A.}~\bibnamefont{Kumar}},
  \bibinfo{journal}{JHEP} \textbf{\bibinfo{volume}{11}}, \bibinfo{pages}{080}
  (\bibinfo{year}{2017}), \eprint{1707.06587}.

\bibitem[{\citenamefont{Borah et~al.}(2019)\citenamefont{Borah, Dev, and
  Kumar}}]{Borah:2018rca}
\bibinfo{author}{\bibfnamefont{D.}~\bibnamefont{Borah}},
  \bibinfo{author}{\bibfnamefont{P.~S.~B.} \bibnamefont{Dev}},
  \bibnamefont{and} \bibinfo{author}{\bibfnamefont{A.}~\bibnamefont{Kumar}},
  \bibinfo{journal}{Phys. Rev.} \textbf{\bibinfo{volume}{D99}},
  \bibinfo{pages}{055012} (\bibinfo{year}{2019}), \eprint{1810.03645}.
  
  
\bibitem{Allahverdi:2007wt} 
  R.~Allahverdi, B.~Dutta and A.~Mazumdar,
  Phys.\ Rev.\ Lett.\  {\bf 99}, 261301 (2007)
  [arXiv:0708.3983 [hep-ph]].
  
\bibitem{Mazumdar:2012qk} 
  A.~Mazumdar and S.~Morisi,
  Phys.\ Rev.\ D {\bf 86}, 045031 (2012)
  [arXiv:1201.6189 [hep-ph]].
  
  
\bibitem{Mazumdar:2010sa} 
  A.~Mazumdar and J.~Rocher,
  Phys.\ Rept.\  {\bf 497}, 85 (2011)
  [arXiv:1001.0993 [hep-ph]].
  
\bibitem{Kohri:2009ka} 
  K.~Kohri, A.~Mazumdar and N.~Sahu,
  Phys.\ Rev.\ D {\bf 80}, 103504 (2009)
  [arXiv:0905.1625 [hep-ph]].
  
 
\bibitem{Rodrigues:2018jpv} 
  J.~G.~Rodrigues, A.~C.~O.~Santos, J.~G.~Ferreira and C.~A.~de S.Pires,
  arXiv:1807.02204 [hep-ph].
  
  
\bibitem{Kazanas:2004kv} 
  D.~Kazanas, R.~N.~Mohapatra, S.~Nasri and V.~L.~Teplitz,
  Phys.\ Rev.\ D {\bf 70}, 033015 (2004)
  [hep-ph/0403291].
  
  

\bibitem[{\citenamefont{Kofman et~al.}(1994)\citenamefont{Kofman, Linde, and
  Starobinsky}}]{Kofman:1994rk}
\bibinfo{author}{\bibfnamefont{L.}~\bibnamefont{Kofman}},
  \bibinfo{author}{\bibfnamefont{A.~D.} \bibnamefont{Linde}}, \bibnamefont{and}
  \bibinfo{author}{\bibfnamefont{A.~A.} \bibnamefont{Starobinsky}},
  \bibinfo{journal}{Phys. Rev. Lett.} \textbf{\bibinfo{volume}{73}},
  \bibinfo{pages}{3195} (\bibinfo{year}{1994}), \eprint{hep-th/9405187}.

\bibitem[{\citenamefont{Kofman et~al.}(1997)\citenamefont{Kofman, Linde, and
  Starobinsky}}]{Kofman:1997yn}
\bibinfo{author}{\bibfnamefont{L.}~\bibnamefont{Kofman}},
  \bibinfo{author}{\bibfnamefont{A.~D.} \bibnamefont{Linde}}, \bibnamefont{and}
  \bibinfo{author}{\bibfnamefont{A.~A.} \bibnamefont{Starobinsky}},
  \bibinfo{journal}{Phys. Rev.} \textbf{\bibinfo{volume}{D56}},
  \bibinfo{pages}{3258} (\bibinfo{year}{1997}), \eprint{hep-ph/9704452}.

\bibitem[{\citenamefont{Liddle and Urena-Lopez}(2006)}]{Liddle:2006qz}
\bibinfo{author}{\bibfnamefont{A.~R.} \bibnamefont{Liddle}} \bibnamefont{and}
  \bibinfo{author}{\bibfnamefont{L.~A.} \bibnamefont{Urena-Lopez}},
  \bibinfo{journal}{Phys. Rev. Lett.} \textbf{\bibinfo{volume}{97}},
  \bibinfo{pages}{161301} (\bibinfo{year}{2006}), \eprint{astro-ph/0605205}.

\bibitem[{\citenamefont{Cardenas}(2007)}]{Cardenas:2007xh}
\bibinfo{author}{\bibfnamefont{V.~H.} \bibnamefont{Cardenas}},
  \bibinfo{journal}{Phys. Rev.} \textbf{\bibinfo{volume}{D75}},
  \bibinfo{pages}{083512} (\bibinfo{year}{2007}), \eprint{astro-ph/0701624}.

\bibitem[{\citenamefont{Panotopoulos}(2007)}]{Panotopoulos:2007ri}
\bibinfo{author}{\bibfnamefont{G.}~\bibnamefont{Panotopoulos}},
  \bibinfo{journal}{Phys. Rev.} \textbf{\bibinfo{volume}{D75}},
  \bibinfo{pages}{127301} (\bibinfo{year}{2007}), \eprint{0706.2237}.

\bibitem[{\citenamefont{Liddle et~al.}(2008)\citenamefont{Liddle, Pahud, and
  Urena-Lopez}}]{Liddle:2008bm}
\bibinfo{author}{\bibfnamefont{A.~R.} \bibnamefont{Liddle}},
  \bibinfo{author}{\bibfnamefont{C.}~\bibnamefont{Pahud}}, \bibnamefont{and}
  \bibinfo{author}{\bibfnamefont{L.~A.} \bibnamefont{Urena-Lopez}},
  \bibinfo{journal}{Phys. Rev.} \textbf{\bibinfo{volume}{D77}},
  \bibinfo{pages}{121301} (\bibinfo{year}{2008}), \eprint{0804.0869}.

\bibitem[{\citenamefont{Bose and Majumdar}(2009)}]{Bose:2009kc}
\bibinfo{author}{\bibfnamefont{N.}~\bibnamefont{Bose}} \bibnamefont{and}
  \bibinfo{author}{\bibfnamefont{A.~S.} \bibnamefont{Majumdar}},
  \bibinfo{journal}{Phys. Rev.} \textbf{\bibinfo{volume}{D80}},
  \bibinfo{pages}{103508} (\bibinfo{year}{2009}), \eprint{0907.2330}.

\bibitem[{\citenamefont{Lerner and McDonald}(2009)}]{Lerner:2009xg}
\bibinfo{author}{\bibfnamefont{R.~N.} \bibnamefont{Lerner}} \bibnamefont{and}
  \bibinfo{author}{\bibfnamefont{J.}~\bibnamefont{McDonald}},
  \bibinfo{journal}{Phys. Rev.} \textbf{\bibinfo{volume}{D80}},
  \bibinfo{pages}{123507} (\bibinfo{year}{2009}), \eprint{0909.0520}.

\bibitem[{\citenamefont{Okada and Shafi}(2011)}]{Okada:2010jd}
\bibinfo{author}{\bibfnamefont{N.}~\bibnamefont{Okada}} \bibnamefont{and}
  \bibinfo{author}{\bibfnamefont{Q.}~\bibnamefont{Shafi}},
  \bibinfo{journal}{Phys. Rev.} \textbf{\bibinfo{volume}{D84}},
  \bibinfo{pages}{043533} (\bibinfo{year}{2011}), \eprint{1007.1672}.

\bibitem[{\citenamefont{De-Santiago and
  Cervantes-Cota}(2011)}]{DeSantiago:2011qb}
\bibinfo{author}{\bibfnamefont{J.}~\bibnamefont{De-Santiago}} \bibnamefont{and}
  \bibinfo{author}{\bibfnamefont{J.~L.} \bibnamefont{Cervantes-Cota}},
  \bibinfo{journal}{Phys. Rev.} \textbf{\bibinfo{volume}{D83}},
  \bibinfo{pages}{063502} (\bibinfo{year}{2011}), \eprint{1102.1777}.

\bibitem[{\citenamefont{Lerner and McDonald}(2011)}]{Lerner:2011ge}
\bibinfo{author}{\bibfnamefont{R.~N.} \bibnamefont{Lerner}} \bibnamefont{and}
  \bibinfo{author}{\bibfnamefont{J.}~\bibnamefont{McDonald}},
  \bibinfo{journal}{Phys. Rev.} \textbf{\bibinfo{volume}{D83}},
  \bibinfo{pages}{123522} (\bibinfo{year}{2011}), \eprint{1104.2468}.

\bibitem[{\citenamefont{de~la Macorra}(2012)}]{delaMacorra:2012sb}
\bibinfo{author}{\bibfnamefont{A.}~\bibnamefont{de~la Macorra}},
  \bibinfo{journal}{Astropart. Phys.} \textbf{\bibinfo{volume}{35}},
  \bibinfo{pages}{478} (\bibinfo{year}{2012}), \eprint{1201.6302}.

\bibitem[{\citenamefont{Khoze}(2013)}]{Khoze:2013uia}
\bibinfo{author}{\bibfnamefont{V.~V.} \bibnamefont{Khoze}},
  \bibinfo{journal}{JHEP} \textbf{\bibinfo{volume}{11}}, \bibinfo{pages}{215}
  (\bibinfo{year}{2013}), \eprint{1308.6338}.

\bibitem[{\citenamefont{Kahlhoefer and McDonald}(2015)}]{Kahlhoefer:2015jma}
\bibinfo{author}{\bibfnamefont{F.}~\bibnamefont{Kahlhoefer}} \bibnamefont{and}
  \bibinfo{author}{\bibfnamefont{J.}~\bibnamefont{McDonald}},
  \bibinfo{journal}{JCAP} \textbf{\bibinfo{volume}{1511}}, \bibinfo{pages}{015}
  (\bibinfo{year}{2015}), \eprint{1507.03600}.

\bibitem[{\citenamefont{Bastero-Gil et~al.}(2016)\citenamefont{Bastero-Gil,
  Cerezo, and Rosa}}]{Bastero-Gil:2015lga}
\bibinfo{author}{\bibfnamefont{M.}~\bibnamefont{Bastero-Gil}},
  \bibinfo{author}{\bibfnamefont{R.}~\bibnamefont{Cerezo}}, \bibnamefont{and}
  \bibinfo{author}{\bibfnamefont{J.~G.} \bibnamefont{Rosa}},
  \bibinfo{journal}{Phys. Rev.} \textbf{\bibinfo{volume}{D93}},
  \bibinfo{pages}{103531} (\bibinfo{year}{2016}), \eprint{1501.05539}.

\bibitem[{\citenamefont{Tenkanen}(2016)}]{Tenkanen:2016twd}
\bibinfo{author}{\bibfnamefont{T.}~\bibnamefont{Tenkanen}},
  \bibinfo{journal}{JHEP} \textbf{\bibinfo{volume}{09}}, \bibinfo{pages}{049}
  (\bibinfo{year}{2016}), \eprint{1607.01379}.

\bibitem[{\citenamefont{Heurtier}(2017)}]{Heurtier:2017nwl}
\bibinfo{author}{\bibfnamefont{L.}~\bibnamefont{Heurtier}},
  \bibinfo{journal}{JHEP} \textbf{\bibinfo{volume}{12}}, \bibinfo{pages}{072}
  (\bibinfo{year}{2017}), \eprint{1707.08999}.

\bibitem[{\citenamefont{Hooper et~al.}(2018)\citenamefont{Hooper, Krnjaic,
  Long, and Mcdermott}}]{Hooper:2018buz}
\bibinfo{author}{\bibfnamefont{D.}~\bibnamefont{Hooper}},
  \bibinfo{author}{\bibfnamefont{G.}~\bibnamefont{Krnjaic}},
  \bibinfo{author}{\bibfnamefont{A.~J.} \bibnamefont{Long}}, \bibnamefont{and}
  \bibinfo{author}{\bibfnamefont{S.~D.} \bibnamefont{Mcdermott}}
  (\bibinfo{year}{2018}), \eprint{1807.03308}.

\bibitem[{\citenamefont{Daido et~al.}(2018)\citenamefont{Daido, Takahashi, and
  Yin}}]{Daido:2017tbr}
\bibinfo{author}{\bibfnamefont{R.}~\bibnamefont{Daido}},
  \bibinfo{author}{\bibfnamefont{F.}~\bibnamefont{Takahashi}},
  \bibnamefont{and} \bibinfo{author}{\bibfnamefont{W.}~\bibnamefont{Yin}},
  \bibinfo{journal}{JHEP} \textbf{\bibinfo{volume}{02}}, \bibinfo{pages}{104}
  (\bibinfo{year}{2018}), \eprint{1710.11107}.

\bibitem[{\citenamefont{Daido et~al.}(2017)\citenamefont{Daido, Takahashi, and
  Yin}}]{Daido:2017wwb}
\bibinfo{author}{\bibfnamefont{R.}~\bibnamefont{Daido}},
  \bibinfo{author}{\bibfnamefont{F.}~\bibnamefont{Takahashi}},
  \bibnamefont{and} \bibinfo{author}{\bibfnamefont{W.}~\bibnamefont{Yin}},
  \bibinfo{journal}{JCAP} \textbf{\bibinfo{volume}{1705}}, \bibinfo{pages}{044}
  (\bibinfo{year}{2017}), \eprint{1702.03284}.

\bibitem[{\citenamefont{Almeida et~al.}(2019)\citenamefont{Almeida, Bernal,
  Rubio, and Tenkanen}}]{Almeida:2018oid}
\bibinfo{author}{\bibfnamefont{J.~P.~B.} \bibnamefont{Almeida}},
  \bibinfo{author}{\bibfnamefont{N.}~\bibnamefont{Bernal}},
  \bibinfo{author}{\bibfnamefont{J.}~\bibnamefont{Rubio}}, \bibnamefont{and}
  \bibinfo{author}{\bibfnamefont{T.}~\bibnamefont{Tenkanen}},
  \bibinfo{journal}{JCAP} \textbf{\bibinfo{volume}{1903}}, \bibinfo{pages}{012}
  (\bibinfo{year}{2019}), \eprint{1811.09640}.

\bibitem[{\citenamefont{Torres~Manso and Rosa}(2019)}]{Manso:2018cba}
\bibinfo{author}{\bibfnamefont{A.}~\bibnamefont{Torres~Manso}}
  \bibnamefont{and} \bibinfo{author}{\bibfnamefont{J.~G.} \bibnamefont{Rosa}},
  \bibinfo{journal}{JHEP} \textbf{\bibinfo{volume}{02}}, \bibinfo{pages}{020}
  (\bibinfo{year}{2019}), \eprint{1811.02302}.

\bibitem[{\citenamefont{Choi et~al.}(2019)\citenamefont{Choi, Kang, Lee, and
  Yamashita}}]{Choi:2019osi}
\bibinfo{author}{\bibfnamefont{S.-M.} \bibnamefont{Choi}},
  \bibinfo{author}{\bibfnamefont{Y.-J.} \bibnamefont{Kang}},
  \bibinfo{author}{\bibfnamefont{H.~M.} \bibnamefont{Lee}}, \bibnamefont{and}
  \bibinfo{author}{\bibfnamefont{K.}~\bibnamefont{Yamashita}}
  (\bibinfo{year}{2019}), \eprint{1902.03781}.

\bibitem[{\citenamefont{Minkowski}(1977)}]{Minkowski:1977sc}
\bibinfo{author}{\bibfnamefont{P.}~\bibnamefont{Minkowski}},
  \bibinfo{journal}{Phys. Lett.} \textbf{\bibinfo{volume}{67B}},
  \bibinfo{pages}{421} (\bibinfo{year}{1977}).

\bibitem[{\citenamefont{Gell-Mann et~al.}(1979)\citenamefont{Gell-Mann, Ramond,
  and Slansky}}]{GellMann:1980vs}
\bibinfo{author}{\bibfnamefont{M.}~\bibnamefont{Gell-Mann}},
  \bibinfo{author}{\bibfnamefont{P.}~\bibnamefont{Ramond}}, \bibnamefont{and}
  \bibinfo{author}{\bibfnamefont{R.}~\bibnamefont{Slansky}},
  \bibinfo{journal}{Conf. Proc.} \textbf{\bibinfo{volume}{C790927}},
  \bibinfo{pages}{315} (\bibinfo{year}{1979}), \eprint{1306.4669}.

\bibitem[{\citenamefont{Mohapatra and Senjanovic}(1980)}]{Mohapatra:1979ia}
\bibinfo{author}{\bibfnamefont{R.~N.} \bibnamefont{Mohapatra}}
  \bibnamefont{and}
  \bibinfo{author}{\bibfnamefont{G.}~\bibnamefont{Senjanovic}},
  \bibinfo{journal}{Phys. Rev. Lett.} \textbf{\bibinfo{volume}{44}},
  \bibinfo{pages}{912} (\bibinfo{year}{1980}).

\bibitem[{\citenamefont{Schechter and Valle}(1980)}]{Schechter:1980gr}
\bibinfo{author}{\bibfnamefont{J.}~\bibnamefont{Schechter}} \bibnamefont{and}
  \bibinfo{author}{\bibfnamefont{J.~W.~F.} \bibnamefont{Valle}},
  \bibinfo{journal}{Phys. Rev.} \textbf{\bibinfo{volume}{D22}},
  \bibinfo{pages}{2227} (\bibinfo{year}{1980}).

\bibitem[{\citenamefont{Babu and He}(1989)}]{Babu:1988yq}
\bibinfo{author}{\bibfnamefont{K.~S.} \bibnamefont{Babu}} \bibnamefont{and}
  \bibinfo{author}{\bibfnamefont{X.~G.} \bibnamefont{He}},
  \bibinfo{journal}{Mod. Phys. Lett.} \textbf{\bibinfo{volume}{A4}},
  \bibinfo{pages}{61} (\bibinfo{year}{1989}).

\bibitem[{\citenamefont{Peltoniemi et~al.}(1993)\citenamefont{Peltoniemi,
  Tommasini, and Valle}}]{Peltoniemi:1992ss}
\bibinfo{author}{\bibfnamefont{J.~T.} \bibnamefont{Peltoniemi}},
  \bibinfo{author}{\bibfnamefont{D.}~\bibnamefont{Tommasini}},
  \bibnamefont{and} \bibinfo{author}{\bibfnamefont{J.~W.~F.}
  \bibnamefont{Valle}}, \bibinfo{journal}{Phys. Lett.}
  \textbf{\bibinfo{volume}{B298}}, \bibinfo{pages}{383} (\bibinfo{year}{1993}).

\bibitem[{\citenamefont{Centelles~Chuliá
  et~al.}(2017{\natexlab{a}})\citenamefont{Centelles~Chulia, Ma, Srivastava,
  and Valle}}]{Chulia:2016ngi}
\bibinfo{author}{\bibfnamefont{S.}~\bibnamefont{Centelles~Chulia}},
  \bibinfo{author}{\bibfnamefont{E.}~\bibnamefont{Ma}},
  \bibinfo{author}{\bibfnamefont{R.}~\bibnamefont{Srivastava}},
  \bibnamefont{and} \bibinfo{author}{\bibfnamefont{J.~W.~F.}
  \bibnamefont{Valle}}, \bibinfo{journal}{Phys. Lett.}
  \textbf{\bibinfo{volume}{B767}}, \bibinfo{pages}{209}
  (\bibinfo{year}{2017}{\natexlab{a}}), \eprint{1606.04543}.

\bibitem[{\citenamefont{Aranda et~al.}(2014)\citenamefont{Aranda, Bonilla,
  Morisi, Peinado, and Valle}}]{Aranda:2013gga}
\bibinfo{author}{\bibfnamefont{A.}~\bibnamefont{Aranda}},
  \bibinfo{author}{\bibfnamefont{C.}~\bibnamefont{Bonilla}},
  \bibinfo{author}{\bibfnamefont{S.}~\bibnamefont{Morisi}},
  \bibinfo{author}{\bibfnamefont{E.}~\bibnamefont{Peinado}}, \bibnamefont{and}
  \bibinfo{author}{\bibfnamefont{J.~W.~F.} \bibnamefont{Valle}},
  \bibinfo{journal}{Phys. Rev.} \textbf{\bibinfo{volume}{D89}},
  \bibinfo{pages}{033001} (\bibinfo{year}{2014}), \eprint{1307.3553}.

\bibitem[{\citenamefont{Chen et~al.}(2016)\citenamefont{Chen, Ding, Rojas,
  Vaquera-Araujo, and Valle}}]{Chen:2015jta}
\bibinfo{author}{\bibfnamefont{P.}~\bibnamefont{Chen}},
  \bibinfo{author}{\bibfnamefont{G.-J.} \bibnamefont{Ding}},
  \bibinfo{author}{\bibfnamefont{A.~D.} \bibnamefont{Rojas}},
  \bibinfo{author}{\bibfnamefont{C.~A.} \bibnamefont{Vaquera-Araujo}},
  \bibnamefont{and} \bibinfo{author}{\bibfnamefont{J.~W.~F.}
  \bibnamefont{Valle}}, \bibinfo{journal}{JHEP} \textbf{\bibinfo{volume}{01}},
  \bibinfo{pages}{007} (\bibinfo{year}{2016}), \eprint{1509.06683}.

\bibitem[{\citenamefont{Ma et~al.}(2015)\citenamefont{Ma, Pollard, Srivastava,
  and Zakeri}}]{Ma:2015mjd}
\bibinfo{author}{\bibfnamefont{E.}~\bibnamefont{Ma}},
  \bibinfo{author}{\bibfnamefont{N.}~\bibnamefont{Pollard}},
  \bibinfo{author}{\bibfnamefont{R.}~\bibnamefont{Srivastava}},
  \bibnamefont{and} \bibinfo{author}{\bibfnamefont{M.}~\bibnamefont{Zakeri}},
  \bibinfo{journal}{Phys. Lett.} \textbf{\bibinfo{volume}{B750}},
  \bibinfo{pages}{135} (\bibinfo{year}{2015}), \eprint{1507.03943}.

\bibitem[{\citenamefont{Reig et~al.}(2016)\citenamefont{Reig, Valle, and
  Vaquera-Araujo}}]{Reig:2016ewy}
\bibinfo{author}{\bibfnamefont{M.}~\bibnamefont{Reig}},
  \bibinfo{author}{\bibfnamefont{J.~W.~F.} \bibnamefont{Valle}},
  \bibnamefont{and} \bibinfo{author}{\bibfnamefont{C.~A.}
  \bibnamefont{Vaquera-Araujo}}, \bibinfo{journal}{Phys. Rev.}
  \textbf{\bibinfo{volume}{D94}}, \bibinfo{pages}{033012}
  (\bibinfo{year}{2016}), \eprint{1606.08499}.

\bibitem[{\citenamefont{Wang and Han}(2016)}]{Wang:2016lve}
\bibinfo{author}{\bibfnamefont{W.}~\bibnamefont{Wang}} \bibnamefont{and}
  \bibinfo{author}{\bibfnamefont{Z.-L.} \bibnamefont{Han}}
  (\bibinfo{year}{2016}), \bibinfo{note}{[JHEP04,166(2017)]},
  \eprint{1611.03240}.

\bibitem[{\citenamefont{Wang et~al.}(2017)\citenamefont{Wang, Wang, Han, and
  Han}}]{Wang:2017mcy}
\bibinfo{author}{\bibfnamefont{W.}~\bibnamefont{Wang}},
  \bibinfo{author}{\bibfnamefont{R.}~\bibnamefont{Wang}},
  \bibinfo{author}{\bibfnamefont{Z.-L.} \bibnamefont{Han}}, \bibnamefont{and}
  \bibinfo{author}{\bibfnamefont{J.-Z.} \bibnamefont{Han}},
  \bibinfo{journal}{Eur. Phys. J.} \textbf{\bibinfo{volume}{C77}},
  \bibinfo{pages}{889} (\bibinfo{year}{2017}), \eprint{1705.00414}.

\bibitem[{\citenamefont{Wang et~al.}(2006)\citenamefont{Wang, Wang, and
  Yang}}]{Wang:2006jy}
\bibinfo{author}{\bibfnamefont{F.}~\bibnamefont{Wang}},
  \bibinfo{author}{\bibfnamefont{W.}~\bibnamefont{Wang}}, \bibnamefont{and}
  \bibinfo{author}{\bibfnamefont{J.~M.} \bibnamefont{Yang}},
  \bibinfo{journal}{Europhys. Lett.} \textbf{\bibinfo{volume}{76}},
  \bibinfo{pages}{388} (\bibinfo{year}{2006}), \eprint{hep-ph/0601018}.

\bibitem[{\citenamefont{Gabriel and Nandi}(2007)}]{Gabriel:2006ns}
\bibinfo{author}{\bibfnamefont{S.}~\bibnamefont{Gabriel}} \bibnamefont{and}
  \bibinfo{author}{\bibfnamefont{S.}~\bibnamefont{Nandi}},
  \bibinfo{journal}{Phys. Lett.} \textbf{\bibinfo{volume}{B655}},
  \bibinfo{pages}{141} (\bibinfo{year}{2007}), \eprint{hep-ph/0610253}.

\bibitem[{\citenamefont{Davidson and Logan}(2009)}]{Davidson:2009ha}
\bibinfo{author}{\bibfnamefont{S.~M.} \bibnamefont{Davidson}} \bibnamefont{and}
  \bibinfo{author}{\bibfnamefont{H.~E.} \bibnamefont{Logan}},
  \bibinfo{journal}{Phys. Rev.} \textbf{\bibinfo{volume}{D80}},
  \bibinfo{pages}{095008} (\bibinfo{year}{2009}), \eprint{0906.3335}.

\bibitem[{\citenamefont{Davidson and Logan}(2010)}]{Davidson:2010sf}
\bibinfo{author}{\bibfnamefont{S.~M.} \bibnamefont{Davidson}} \bibnamefont{and}
  \bibinfo{author}{\bibfnamefont{H.~E.} \bibnamefont{Logan}},
  \bibinfo{journal}{Phys. Rev.} \textbf{\bibinfo{volume}{D82}},
  \bibinfo{pages}{115031} (\bibinfo{year}{2010}), \eprint{1009.4413}.

\bibitem[{\citenamefont{Bonilla and Valle}(2016)}]{Bonilla:2016zef}
\bibinfo{author}{\bibfnamefont{C.}~\bibnamefont{Bonilla}} \bibnamefont{and}
  \bibinfo{author}{\bibfnamefont{J.~W.~F.} \bibnamefont{Valle}},
  \bibinfo{journal}{Phys. Lett.} \textbf{\bibinfo{volume}{B762}},
  \bibinfo{pages}{162} (\bibinfo{year}{2016}), \eprint{1605.08362}.

\bibitem[{\citenamefont{Farzan and Ma}(2012)}]{Farzan:2012sa}
\bibinfo{author}{\bibfnamefont{Y.}~\bibnamefont{Farzan}} \bibnamefont{and}
  \bibinfo{author}{\bibfnamefont{E.}~\bibnamefont{Ma}}, \bibinfo{journal}{Phys.
  Rev.} \textbf{\bibinfo{volume}{D86}}, \bibinfo{pages}{033007}
  (\bibinfo{year}{2012}), \eprint{1204.4890}.

\bibitem[{\citenamefont{Bonilla et~al.}(2016)\citenamefont{Bonilla, Ma,
  Peinado, and Valle}}]{Bonilla:2016diq}
\bibinfo{author}{\bibfnamefont{C.}~\bibnamefont{Bonilla}},
  \bibinfo{author}{\bibfnamefont{E.}~\bibnamefont{Ma}},
  \bibinfo{author}{\bibfnamefont{E.}~\bibnamefont{Peinado}}, \bibnamefont{and}
  \bibinfo{author}{\bibfnamefont{J.~W.~F.} \bibnamefont{Valle}},
  \bibinfo{journal}{Phys. Lett.} \textbf{\bibinfo{volume}{B762}},
  \bibinfo{pages}{214} (\bibinfo{year}{2016}), \eprint{1607.03931}.

\bibitem[{\citenamefont{Ma and Popov}(2017)}]{Ma:2016mwh}
\bibinfo{author}{\bibfnamefont{E.}~\bibnamefont{Ma}} \bibnamefont{and}
  \bibinfo{author}{\bibfnamefont{O.}~\bibnamefont{Popov}},
  \bibinfo{journal}{Phys. Lett.} \textbf{\bibinfo{volume}{B764}},
  \bibinfo{pages}{142} (\bibinfo{year}{2017}), \eprint{1609.02538}.

\bibitem[{\citenamefont{Ma and Sarkar}(2018)}]{Ma:2017kgb}
\bibinfo{author}{\bibfnamefont{E.}~\bibnamefont{Ma}} \bibnamefont{and}
  \bibinfo{author}{\bibfnamefont{U.}~\bibnamefont{Sarkar}},
  \bibinfo{journal}{Phys. Lett.} \textbf{\bibinfo{volume}{B776}},
  \bibinfo{pages}{54} (\bibinfo{year}{2018}), \eprint{1707.07698}.

\bibitem[{\citenamefont{Borah}(2016)}]{Borah:2016lrl}
\bibinfo{author}{\bibfnamefont{D.}~\bibnamefont{Borah}},
  \bibinfo{journal}{Phys. Rev.} \textbf{\bibinfo{volume}{D94}},
  \bibinfo{pages}{075024} (\bibinfo{year}{2016}), \eprint{1607.00244}.

\bibitem[{\citenamefont{Borah and Dasgupta}(2016)}]{Borah:2016zbd}
\bibinfo{author}{\bibfnamefont{D.}~\bibnamefont{Borah}} \bibnamefont{and}
  \bibinfo{author}{\bibfnamefont{A.}~\bibnamefont{Dasgupta}},
  \bibinfo{journal}{JCAP} \textbf{\bibinfo{volume}{1612}}, \bibinfo{pages}{034}
  (\bibinfo{year}{2016}), \eprint{1608.03872}.

\bibitem[{\citenamefont{Borah and
  Dasgupta}(2017{\natexlab{a}})}]{Borah:2016hqn}
\bibinfo{author}{\bibfnamefont{D.}~\bibnamefont{Borah}} \bibnamefont{and}
  \bibinfo{author}{\bibfnamefont{A.}~\bibnamefont{Dasgupta}},
  \bibinfo{journal}{JHEP} \textbf{\bibinfo{volume}{01}}, \bibinfo{pages}{072}
  (\bibinfo{year}{2017}{\natexlab{a}}), \eprint{1609.04236}.

\bibitem[{\citenamefont{Borah and
  Dasgupta}(2017{\natexlab{b}})}]{Borah:2017leo}
\bibinfo{author}{\bibfnamefont{D.}~\bibnamefont{Borah}} \bibnamefont{and}
  \bibinfo{author}{\bibfnamefont{A.}~\bibnamefont{Dasgupta}},
  \bibinfo{journal}{JCAP} \textbf{\bibinfo{volume}{1706}}, \bibinfo{pages}{003}
  (\bibinfo{year}{2017}{\natexlab{b}}), \eprint{1702.02877}.

\bibitem[{\citenamefont{Centelles~Chuliá
  et~al.}(2017{\natexlab{b}})\citenamefont{Centelles~Chulia, Srivastava, and
  Valle}}]{CentellesChulia:2017koy}
\bibinfo{author}{\bibfnamefont{S.}~\bibnamefont{Centelles~Chulia}},
  \bibinfo{author}{\bibfnamefont{R.}~\bibnamefont{Srivastava}},
  \bibnamefont{and} \bibinfo{author}{\bibfnamefont{J.~W.~F.}
  \bibnamefont{Valle}}, \bibinfo{journal}{Phys. Lett.}
  \textbf{\bibinfo{volume}{B773}}, \bibinfo{pages}{26}
  (\bibinfo{year}{2017}{\natexlab{b}}), \eprint{1706.00210}.

\bibitem[{\citenamefont{Bonilla et~al.}(2018)\citenamefont{Bonilla, Lamprea,
  Peinado, and Valle}}]{Bonilla:2017ekt}
\bibinfo{author}{\bibfnamefont{C.}~\bibnamefont{Bonilla}},
  \bibinfo{author}{\bibfnamefont{J.~M.} \bibnamefont{Lamprea}},
  \bibinfo{author}{\bibfnamefont{E.}~\bibnamefont{Peinado}}, \bibnamefont{and}
  \bibinfo{author}{\bibfnamefont{J.~W.~F.} \bibnamefont{Valle}},
  \bibinfo{journal}{Phys. Lett.} \textbf{\bibinfo{volume}{B779}},
  \bibinfo{pages}{257} (\bibinfo{year}{2018}), \eprint{1710.06498}.

\bibitem[{\citenamefont{Memenga et~al.}(2013)\citenamefont{Memenga, Rodejohann,
  and Zhang}}]{Memenga:2013vc}
\bibinfo{author}{\bibfnamefont{N.}~\bibnamefont{Memenga}},
  \bibinfo{author}{\bibfnamefont{W.}~\bibnamefont{Rodejohann}},
  \bibnamefont{and} \bibinfo{author}{\bibfnamefont{H.}~\bibnamefont{Zhang}},
  \bibinfo{journal}{Phys. Rev.} \textbf{\bibinfo{volume}{D87}},
  \bibinfo{pages}{053021} (\bibinfo{year}{2013}), \eprint{1301.2963}.



\bibitem[{\citenamefont{Centelles~Chuliá
  et~al.}(2018{\natexlab{a}})\citenamefont{Centelles~Chulia, Srivastava, and
  Valle}}]{CentellesChulia:2018gwr}
\bibinfo{author}{\bibfnamefont{S.}~\bibnamefont{Centelles~Chulia}},
  \bibinfo{author}{\bibfnamefont{R.}~\bibnamefont{Srivastava}},
  \bibnamefont{and} \bibinfo{author}{\bibfnamefont{J.~W.~F.}
  \bibnamefont{Valle}}, \bibinfo{journal}{Phys. Lett.}
  \textbf{\bibinfo{volume}{B781}}, \bibinfo{pages}{122}
  (\bibinfo{year}{2018}{\natexlab{a}}), \eprint{1802.05722}.

\bibitem[{\citenamefont{Centelles~Chuliá
  et~al.}(2018{\natexlab{b}})\citenamefont{Centelles~Chulia, Srivastava, and
  Valle}}]{CentellesChulia:2018bkz}
\bibinfo{author}{\bibfnamefont{S.}~\bibnamefont{Centelles~Chulia}},
  \bibinfo{author}{\bibfnamefont{R.}~\bibnamefont{Srivastava}},
  \bibnamefont{and} \bibinfo{author}{\bibfnamefont{J.~W.~F.}
  \bibnamefont{Valle}} (\bibinfo{year}{2018}{\natexlab{b}}),
  \eprint{1804.03181}.

\bibitem[{\citenamefont{Han and Wang}(2018)}]{Han:2018zcn}
\bibinfo{author}{\bibfnamefont{Z.-L.} \bibnamefont{Han}} \bibnamefont{and}
  \bibinfo{author}{\bibfnamefont{W.}~\bibnamefont{Wang}}
  (\bibinfo{year}{2018}), \eprint{1805.02025}.

\bibitem[{\citenamefont{Borah et~al.}(2018)\citenamefont{Borah, Karmakar, and
  Nanda}}]{Borah:2018gjk}
\bibinfo{author}{\bibfnamefont{D.}~\bibnamefont{Borah}},
  \bibinfo{author}{\bibfnamefont{B.}~\bibnamefont{Karmakar}}, \bibnamefont{and}
  \bibinfo{author}{\bibfnamefont{D.}~\bibnamefont{Nanda}},
  \bibinfo{journal}{JCAP} \textbf{\bibinfo{volume}{1807}}, \bibinfo{pages}{039}
  (\bibinfo{year}{2018}), \eprint{1805.11115}.
  
  
\bibitem[{\citenamefont{Borah and Karmakar}(2018)}]{Borah:2017dmk}
\bibinfo{author}{\bibfnamefont{D.}~\bibnamefont{Borah}} \bibnamefont{and}
  \bibinfo{author}{\bibfnamefont{B.}~\bibnamefont{Karmakar}},
  \bibinfo{journal}{Phys. Lett.} \textbf{\bibinfo{volume}{B780}},
  \bibinfo{pages}{461} (\bibinfo{year}{2018}), \eprint{1712.06407}.
  
    
\bibitem[{\citenamefont{Borah and Karmakar}(2019)}]{Borah:2018nvu}
\bibinfo{author}{\bibfnamefont{D.}~\bibnamefont{Borah}} \bibnamefont{and}
  \bibinfo{author}{\bibfnamefont{B.}~\bibnamefont{Karmakar}},
  \bibinfo{journal}{Phys. Lett.} \textbf{\bibinfo{volume}{B789}},
  \bibinfo{pages}{59} (\bibinfo{year}{2019}), \eprint{1806.10685}.

   \bibitem{Langacker:2008yv}
  P. Langacker, Rev. Mod. Phys. {\bf 81}, 1199 (2009)
  [arXiv:0801.1345 [hep-ph]].
  
  \bibitem{Nanda:2019nqy}
  D. Nanda and D. Borah, [arXiv:1911.04703 [hep-ph]].
  
  
\bibitem{Dong:2010in} 
  X.~Dong, B.~Horn, E.~Silverstein and A.~Westphal,
  Phys.\ Rev.\ D {\bf 84}, 026011 (2011), 
  1011.4521 

  
\bibitem{Evans:2015mta} 
  J.~L.~Evans, T.~Gherghetta and M.~Peloso,
  Phys.\ Rev.\ D {\bf 92}, no. 2, 021303 (2015), 
  1501.06560
  
\bibitem{McAllister:2014mpa} 
  L.~McAllister, E.~Silverstein, A.~Westphal and T.~Wrase,
  JHEP {\bf 1409}, 123 (2014), 
  1405.3652.
  
\bibitem{Buchmuller:2015oma} 
  W.~Buchmuller, E.~Dudas, L.~Heurtier, A.~Westphal, C.~Wieck and M.~W.~Winkler,
  JHEP {\bf 1504}, 058 (2015), 
  1501.05812.

  
\bibitem{Liddle:2003as} 
  A.~R.~Liddle and S.~M.~Leach,
  Phys.\ Rev.\ D {\bf 68}, 103503 (2003)
  [astro-ph/0305263].
  
\bibitem{Martin:2010kz} 
  J.~Martin and C.~Ringeval,
  Phys.\ Rev.\ D {\bf 82}, 023511 (2010)
  [arXiv:1004.5525 [astro-ph.CO]].
  
\bibitem{Dodelson:2003vq} 
  S.~Dodelson and L.~Hui,
  Phys.\ Rev.\ Lett.\  {\bf 91}, 131301 (2003)
  [astro-ph/0305113].
  
\bibitem{Kolb:1990vq} 
  E.~W.~Kolb and M.~S.~Turner,
  Front.\ Phys.\  {\bf 69}, 1 (1990).
  




  \bibitem{Ballesteros:2015iua} 
  G.~Ballesteros and C.~Tamarit,
  JHEP {\bf 1509}, 210 (2015)
  [arXiv:1505.07476 [hep-ph]].
  
  \bibitem[{\citenamefont{Lebedev and Westphal}(2013)}]{Lebedev:2012sy}
\bibinfo{author}{\bibfnamefont{O.}~\bibnamefont{Lebedev}} \bibnamefont{and}
  \bibinfo{author}{\bibfnamefont{A.}~\bibnamefont{Westphal}},
  \bibinfo{journal}{Phys. Lett.} \textbf{\bibinfo{volume}{B719}},
  \bibinfo{pages}{415} (\bibinfo{year}{2013}), \eprint{1210.6987}.
  
  
  
  
\bibitem{Felder:1999pv1} 
  G.~N.~Felder, L.~Kofman and A.~D.~Linde,
  Phys.\ Rev.\ D {\bf 60}, 103505 (1999)
  [hep-ph/9903350].
  
  
  \bibitem{ArmendarizPicon:2007iv} 
  C.~Armendariz-Picon, M.~Trodden and E.~J.~West,
  JCAP {\bf 0804}, 036 (2008)
  [arXiv:0707.2177 [hep-ph]].
  
  
  \bibitem{Desroche:2005yt} 
  M.~Desroche, G.~N.~Felder, J.~M.~Kratochvil and A.~D.~Linde,
  Phys.\ Rev.\ D {\bf 71}, 103516 (2005)
  [hep-th/0501080].
  
  \bibitem{Allahverdi:2011aj} 
  R.~Allahverdi, A.~Ferrantelli, J.~Garcia-Bellido and A.~Mazumdar,
  Phys.\ Rev.\ D {\bf 83}, 123507 (2011)
  [arXiv:1103.2123 [hep-ph]].
  
\bibitem[{\citenamefont{Elahi et~al.}(2015)\citenamefont{Elahi, Kolda, and
  Unwin}}]{Elahi:2014fsa}
\bibinfo{author}{\bibfnamefont{F.}~\bibnamefont{Elahi}},
  \bibinfo{author}{\bibfnamefont{C.}~\bibnamefont{Kolda}}, \bibnamefont{and}
  \bibinfo{author}{\bibfnamefont{J.}~\bibnamefont{Unwin}},
  \bibinfo{journal}{JHEP} \textbf{\bibinfo{volume}{03}}, \bibinfo{pages}{048}
  (\bibinfo{year}{2015}), \eprint{1410.6157}.

\bibitem[{\citenamefont{Feng et~al.}(2003)\citenamefont{Feng, Rajaraman, and
  Takayama}}]{Feng:2003uy}
\bibinfo{author}{\bibfnamefont{J.~L.} \bibnamefont{Feng}},
  \bibinfo{author}{\bibfnamefont{A.}~\bibnamefont{Rajaraman}},
  \bibnamefont{and} \bibinfo{author}{\bibfnamefont{F.}~\bibnamefont{Takayama}},
  \bibinfo{journal}{Phys. Rev.} \textbf{\bibinfo{volume}{D68}},
  \bibinfo{pages}{063504} (\bibinfo{year}{2003}), \eprint{hep-ph/0306024}.

\bibitem{Coleman:1977py} 
  S.~R.~Coleman,
  Phys.\ Rev.\ D {\bf 15}, 2929 (1977)
  Erratum: [Phys.\ Rev.\ D {\bf 16}, 1248 (1977)].
  
\bibitem{Sher:1988mj} 
  M.~Sher,
  Phys.\ Rept.\  {\bf 179}, 273 (1989).

  
\bibitem{Lee:1985uv} 
  K.~M.~Lee and E.~J.~Weinberg,
  Nucl.\ Phys.\ B {\bf 267}, 181 (1986).
  
\bibitem{Arnold:1989cb} 
  P.~B.~Arnold,
  Phys.\ Rev.\ D {\bf 40}, 613 (1989).
\end{thebibliography}

\end{document}